\documentclass{aastex}
\usepackage{epsfig,psfig,fancyheadings}
\usepackage{rotating,longtable}

\shorttitle{Fornax Dwarf Spheroidal Galaxy}
\shortauthors{Coleman et al.}

\begin{document}

\title{A Wide-Field Survey of the Fornax Dwarf Spheroidal Galaxy}

\author{Matthew Coleman}
\affil{Research School of Astronomy \& Astrophysics, Institute of 
Advanced Studies, The Australian National University, Cotter Road, Weston 
Creek, ACT 2611, Australia}
\email{coleman@mso.anu.edu.au}

\author{G. S. Da Costa}
\affil{Research School of Astronomy \& Astrophysics, Institute of 
Advanced Studies, The Australian National University, Cotter Road, Weston 
Creek, ACT 2611, Australia}
\email{gdc@mso.anu.edu.au}

\author{Joss Bland-Hawthorn}
\affil{Anglo-Australian Observatory, PO Box 296, Epping, NSW 2121, Australia}
\email{jbh@aaoepp.aao.gov.au}

\and

\author{Kenneth C. Freeman}
\affil{Research School of Astronomy \& Astrophysics, Institute of 
Advanced Studies, The Australian National University, Cotter Road, Weston 
Creek, ACT 2611, Australia}
\email{kcf@mso.anu.edu.au}

\slugcomment{Version of 9 December, 2004}
\begin{abstract}
The results of a photometric survey of the Fornax dwarf spheroidal galaxy are presented herein.  Accurate photometry in two colours ($V$ and $I$) was collected over a 10 deg${}^2$ area centred on the Fornax system with the aim of searching for extra-tidal structure.  The data were complete to a magnitude of $V=20$, or approximately one magnitude brighter than the Fornax red clump stars.  Stars were selected with a colour and magnitude lying near the Fornax red giant branch, thereby reducing contamination from the field star population.  We were thus able to probe the outer structure of this dwarf galaxy.  Immediately visible was a shell-like structure located 1.3 degrees NW from the centre of Fornax, approximately 30 arcmin beyond the nominal tidal radius at this position angle.  We have measured the absolute visual magnitude of this feature to be $M_V \sim -7$.  The feature is aligned with a previously reported shell (age $\sim$ 2 Gyr) located near the core radius of Fornax.  A statistical analysis of the extra-tidal stellar distribution further revealed two lobes situated on the Fornax minor axis which are aligned with the two shell-like features.  The two-lobed structure combined with the two shells provide strong evidence that Fornax has experienced a merger event in the relatively recent past.
\end{abstract}

\keywords{galaxies: dwarf --- galaxies: individual (Fornax) --- galaxies:
photometry --- galaxies: stellar content --- galaxies: interactions --- 
Galaxy: halo ---  Local Group}

\section{Introduction}
The past two decades has witnessed a revolution in the study of faint galaxies.  With the advent of larger telescopes, more efficient detectors and new observing techniques, we are now able to reach stars down to fainter limiting magnitudes over large areas of the sky.  In recent years, the complex nature of dwarf galaxies has consequently become apparent.  Historically, dwarf spheroidal (dSph) galaxies were seen essentially as large, diffuse globular clusters, uncomplicated structures consisting of a single-age stellar population.  This view has been changed on several fronts.  In terms of star formation, \citet{mould83} first showed that the stellar population of the Carina dSph is complex.  Subsequently, the star formation histories of nearby dwarf spheroidal galaxies have been revealed to be complex and varying from system to system.  For example, whereas the Ursa Minor dSph consists mainly of old stars \citep{olsz85,mighell99,carrera02}, $80\%$ of the stars in Leo I formed in the last seven billion years \citep{gallart99}.

The importance of dwarf galaxies in the dark matter regime has also become apparent.  Measurements of individual stellar velocities in the dSph systems reveal that they are surprisingly massive for their luminosity.  The optical mass-to-light ratios of some dwarf galaxies are found to be $\sim$$100$ or more, indicating that these objects are dominated by dark matter.  Indeed, dwarf galaxies are the darkest objects in the universe which we can directly observe.  The velocity profiles of these objects are flat (for example, \citealt{mateo97,m98,kleyna01}), revealing that the dark halos of dwarf galaxies extend beyond the detected limits of the luminous material.  That is, the luminous material of dwarf galaxies is only a minor component of such objects.

In the standard Cold Dark Matter (CDM) scenario, fluctuations in the power spectrum during the early times of the Universe resulted in small clumps of dark matter with masses $\sim$$10^6 M_{\odot}$ (for example, \citealt{susa04} and references therein).  By redshift $z \sim 10$, dwarf galaxies have formed through the merger of these small dark halos.  Indeed, \citet{ricotti04} predict that despite existing for several crossing times, the inner region of isolated dark halos should exhibit little structural evolution after $z = 10$.  The majority of these dwarf galaxies were not sufficiently isolated to survive to the present day, and have since merged to form the large scale structure now visible.  Therefore, the dwarf galaxies currently orbiting the Galaxy are survivors from this merger process, and are thought to be only a small part of the original low-mass galaxy population.

Hence, dwarf galaxies provide the link between smaller and larger structures.  If dwarf galaxies are built up through the accretion of smaller dark halos, it may be possible to find the primordial building blocks in the local universe \citep{freeman02}.  \citet{kleyna98,kleyna03} claim to have found an relic from the epoch of dwarf galaxy formation in the nearby satellite in Ursa Minor (UMi) in the form of a dense, kinematically cold population of stars, although the presence of DM in UMi has not been categorically proven \citep{spick03}.  Assuming these stars trace the distribution of underlying dark material, then this clump represents a density spike in the distribution of dark matter outside the centre of UMi.  However, although dwarf galaxies are considered to have formed through the merger of smaller systems, such interactions are generally thought to be confined to the early epoch of the Universe.  In contrast, a clump of stars detected in Fornax \citep{coleman04a} was interpreted to be the remnant of the merger between two dSph-sized objects as little as 2 Gyr ago.  This may be related to the enhanced period of star formation found by \citet{pont04}, occuring approximately $2-4$ Gyr ago in concert with a period of increased metal enrichment.  Hence, Fornax could be an important test object for recent dwarf galaxy interactions.

Measuring structural parameters is difficult for nearby dwarf galaxies, since relatively deep, wide-field data are required for these extended, low surface brightness objects.  However, the structure of satellite systems may be used to examine properties of the Galaxy.  Simulations \citep{helmi01,mayer01} predict that the structure of these satellites will be distorted by tidal forces as they orbit the host object.  An extreme form of such structural distortion is evident in the Sagittarius dwarf galaxy \citep{ibata94,majewski03} and the globular cluster Palomar 5 \citep{odenkirchen01,odenkirchen03}.  These two objects have lost a significant portion of their stars to tidal tails caused by the Galactic tidal field.  Several authors have shown that the shape and kinematics of these tails can be used to trace the distribution of dark matter throughout both the host and satellite \citep{sackett94,johnston99,ibata01}.  Also, as the host tidal field heats the satellite, its survivability is highly dependent on its dark matter distribution (for example, see \citealt{majewski03} and references therein).  Therefore, although Fornax is situated far from the Galaxy, it may be possible to detect the first signs of heating due to the Galactic tidal field by studying the outer structure of this dSph.

In this paper, we describe a wide-field survey of the Fornax dSph galaxy.  The aim of this investigation was to search for substructure in Fornax, especially in the poorly-mapped outer regions.  Our goal was to determine if Fornax has a recent merger history, and relate this to its most recent star formation history, and to examine whether the Galactic tidal field is heating the outer regions of this dSph.  A short description of some of these results is given in \citet{coleman05}; here we present a comprehensive analysis.  Section 2 describes previous studies of Fornax, and summarises its stellar population, structure and kinematics.  Section 3 details the Fornax survey, describing the observations and data reduction.  The photometry and calibration is detailed in \S 4, and \S 5 examines the results of the survey.  An analysis of the outer and inner region of Fornax is presented in \S \S 6 and 7 respectively, and we conclude in \S \S 8 and 9.

\section{Fornax} \label{fornax}
The Fornax dSph galaxy was first detected on Harvard blue photographic plates as a diffuse system which could be separated into individual stars only by the most powerful telescopes of the time \citep{shapley39}.  With an integrated $V$-band magnitude of $M_V = -13.2$ and a mass (including dark matter) calculated from the best fit King profile to be $6.8 \times 10^7 M_{\odot}$ \citep{m98}, Fornax is the most massive known object of this type orbiting the Galaxy, excluding the disrupting Sagittarius dSph.  The integrated absolute magnitude of Fornax makes it more luminous than all the other eight known Galactic dSphs put together (excluding Sagittarius).  Fornax covers a large area on the sky, with a tidal radius of $71' \pm 4'$ \citep{ih95}.  It is situated at a distance of $138 \pm 8$ kpc \citep{m98}, and hence the linear tidal radius is $2.85 \pm 0.16$ kpc.

\subsection{Stellar Population and Chemical Enrichment}
Dwarf galaxies are typically small, diffuse systems, yet can contain complex star formation histories.  The star formation history of Fornax appears to be unusually complex in comparison to most other dwarf galaxies.  \citet{demers79} first detected an extended giant branch in the colour-magnitude diagram (CMD), possibly made up of carbon stars.  These were later spectroscopically confirmed by \citet{aaron80,aaron85}.  More recent deep studies have covered a large area of Fornax and found the intermediate-age ($2-8$ Gyr) stars are the dominant population.  \citet{stetson98} covered an area approximately 1/3 deg${}^2$ down to a depth of $V \approx 24$, revealing the intricate history of star formation in Fornax.  This, combined with a more recent study by \citet{saviane00}, indicated that the stars in Fornax can be divided into three main classes: (i) an old population (age $> 10$ Gyr) identified by the old horizontal branch, the old red giant branch (RGB) and RR Lyrae stars; (ii) an intermediate-age population, represented by the upper asymptotic giant branch (AGB), part of the RGB, and the red clump, and; (iii) a population of young (age $\le 2$ Gyr) main sequence stars.

The position of a star in the RGB reflects its age and metallicity, although these quantities exhibit a degeneracy in this region of the CMD.  Therefore, using the position and width of the RGB can lead to erroneous estimates of the metallicity (or metallicity spread) in a system possessing a substantial intermediate age population such as Fornax.  This degeneracy can be broken by collecting spectra for RGB stars, which allows an independent determination of the mean abundance (and its dispersion) in a stellar system.  Once these parameters are known for a large sample of stars, the chemical enrichment history of the stellar system may be pieced together.  Consequently, \citet{tolstoy01} collected spectra for 32 stars in Fornax and determined the mean heavy-metal abundance from the Ca {\sc ii} triplet to be $\langle$[Fe/H]$\rangle = -1.0 \pm 0.3$, with the bulk of Fornax stars between $-1.5$ and $-0.7$.  \citet{pont04} repeated this experiment with 117 red giants towards the centre of Fornax, finding evidence of metal-rich population ($-0.7 \lesssim$ [Fe/H] $\lesssim -0.4$) with ages $\sim 2$ Gyr.  They conclude that the age-metallicity relation for Fornax is complex, and that the system has experienced a significant period of metal enrichment in the last few Gyr.  \citet{pont04} also find that the chemical enrichment history of Fornax makes it more comparable to the LMC or the Galactic disk rather than the remainder of the Galactic dSph population.

In summary, the star formation of Fornax appears to have begun approximately 12 Gyr ago and has continued almost to the present day.  Saviane et al.\ (2000) estimate stars in the blue main sequence may be as young as 200 Myr.  \citet{pont04} find that the metal abundance rose to [Fe/H] $\sim -1.0$ within the first few gigayears, and that the last $\sim$4 Gyr have seen a significant increase in chemical enrichment.  While the formation of stars in Fornax appears to have been a relatively continuous process \citep{tolstoy03}, \citet{buon99} find evidence for gaps in the sub-giant branch, possibly indicating breaks in the star formation.  Indeed, the well-populated subgiant branch indicates a significant population with an age of $2-4$ Gyr.  Stetson et al.\ (1998) observed that the recent star formation in Fornax occurred in the central regions.  However, \citet{young99} searched the same area and was unable to detect a corresponding supply of neutral hydrogen, inferring that the burst of star formation has ionised the gas or possibly ejected it beyond their search area.  An H{\sc i} survey of the Fornax region over a wider field did not detect a significant amount of neutral hydrogen (Bouchard 2004, personal communication).

The various stellar populations show different spatial distributions throughout Fornax.  The old and intermediate populations, represented by the RR Lyrae variables and red clump stars respectively, have the most extended distribution.  Conversely, the young main sequence stars are more centrally concentrated, appearing to define a bar with the position angle of the major axis offset by approximately $30^{\circ}$ from that of the dwarf galaxy.  There is also the possibility that the young stars are concentrated towards both ends of this bar, suggesting a ``bilateral rather than a rotationally symmetric structure'' (Stetson et al.\ 1998).  Thus, recent star formation has occurred irregularly throughout Fornax.  

In addition, first results from the current program \citep{coleman04a} indicated an apparent association of young stars located approximately $17'$ from the centre of Fornax, using the data of Stetson et al.\ (1998).  This feature is aligned with the major axis, located on the minor axis and appears shell-like in nature.  Our inference was that it represented shell structure in Fornax.  If true, this is the first such discovery in a dwarf galaxy, and probably represents a merger between Fornax and another gas-rich companion approximately 2 Gyr ago.  In such an interaction, perturbation of the neutral hydrogen may well lead to an accelerated star formation rate consistent with that observed by \citet{pont04}.

\subsection{Structure}
The first study of the structure of Fornax was by \citet{hodge61b} using star counts from photographic plates.  Hodge constructed a radial profile and contour plot for Fornax, noting two main points concerning its structure: (i) the ellipticity increased with increasing radius, and; (ii) an asymmetry, such that the point of highest density was offset from the centre by $\sim$$6'$.  Two later studies \citep{dev68,hodge74} used photoelectric surface photometry to map the structure of Fornax.  Both confirmed the asymmetry found by \citet{hodge61b}, however these studies were hindered by the faint surface brightness of Fornax.

More recent analyses \citep{eskridge88a,eskridge88b,demers94,ih95} have used wide-field photographic plate data to search for substructure and extra-tidal stars in Fornax.  Although these studies have confirmed the asymmetry in the inner regions ($r < 30'$), the presence of otherwise extra-tidal stars remains controversial.  \citet{ih95} estimate their detection of Fornax stars outside the nominal tidal radius at the $5\sigma$ significance level.  A recent CCD-based survey by \citet{walcher03} covered an area $8.5$ deg${}^2$ around Fornax in the $V$-band.  They did not find evidence of extra-tidal stars, however they claim a significantly increased tidal radius of $98'$.  Hence, the possible tidal distortion of Fornax by the Galaxy is still to be constrained.

\subsection{Kinematics and Dark Matter}
The kinematics of Fornax may provide some insight into the structural development of this dSph.  However, the low heliocentric velocity ($53$ km s${}^{-1}$; \citealt{m98}) makes Fornax unsuitable for large-scale intermediate-resolution spectroscopy with the aim of measuring radial velocities.  In addition, high S/N spectra of a large number of stars in Fornax are difficult to obtain due to its relatively large distance of 138 kpc \citep{m98}.  Consequently, dynamical studies of Fornax have been limited to an analysis of upper RGB stellar velocities.  \citet{mateo91} collected echelle spectroscopy of 44 stars located in two fields of Fornax; one located at the centre and the other approximately two core radii SW along the major axis from the centre of Fornax.  Rejecting some stars based on radial velocities and photometry, they found the central velocity dispersion of Fornax to be $\sim 10$ km s${}^{-1}$.  These results imply a mass-to-light ratio of at least 5.3, and possibly as high as 26.  The large range in these values is a result of the range of published central surface brightness values \citep{mateo91}.

Hence, Fornax appears to contain a significant amount of dark matter; \citet{mateo91} estimate at least $60\%$ of the {\em central} mass of Fornax is dark.  The analysis also found the velocity dispersion of the outer field to be similar to that of the central field, suggesting a flat velocity dispersion profile for the luminous material.  This is unlike a \citet{king66} model; a tidally truncated isotropic system in which mass follows light has a declining velocity dispersion with increasing radius.  Consequently, the dark halo appears to be more extended than the visible component, and may dominate the dynamics of Fornax at all scales.

\subsection{This Survey}
The structural and dynamical history of Fornax is still to be fully determined.  Previous studies have found evidence of substructure in the inner regions, and possible Fornax stars lying beyond the nominal boundary of the system.  These may be indications of the perturbation suffered by Fornax under the gravitational influence of the Galaxy.  However, the attempt to find Fornax structure outside the nominal tidal radius has been limited by the low surface brightness of the region.  At radii approaching the tidal radius the star counts fall into the noise of the background sources.

An effective method to increase the signal-to-noise of stellar counts in the outer regions of nearby systems is to use the CMD to detect candidate extra-tidal stars \citep{grill95}.  We refer the reader to \citet{mart-del01,piatek01,majewski02} for recent examples of this process.  We have compiled a multi-colour spatial map of Fornax using a mosaic of CCD observations.  The inner regions provided a CMD of Fornax, which was used to select candidate Fornax stars beyond its nominal tidal boundary, thereby reducing contamination from background and foreground sources, and resulting in a better model of the outer structure of the object.  This is the first such survey of the Fornax dSph galaxy.

\section{Observations and Data Reduction}
The Fornax dSph covers a relatively large area on the sky, with a major axis length of approximately $2.5^{\circ}$.  Hence, a complete map of this object requires a survey region several degrees across.  CCD images were obtained with the Siding Spring Observatory 1 metre telescope using the Wide-Field Imager (WFI), which comprises eight 4096 $\times$ 2048 CCDs arranged in a $2 \times 4$ mosaic to give a total format of 8192 $\times$ 8192 pixels.  This gives a field-of-view of $52 \times 52$ arcmin${}^2$ at a scale of $0.38''$ per pixel, ideal for covering large areas of the sky to a reasonable depth.  A $4 \times 4$ grid of fields centred on Fornax was observed over four observation runs between October 2001 and December 2003 (Table \ref{fornaxobservations}), covering an area of $3.1^{\circ} \times 3.1^{\circ}$ on the sky.  A $10\%$ overlap region between adjacent fields was used to determine the internal astrometric precision and to ensure the photometric zeropoint was constant across the survey.  A map of these fields is given in Fig.\ \ref{obsmap}, where the sixteen fields are labelled $\mbox{F1}, \mbox{F2}, \dots, \mbox{F16}$.

Each field was observed in the $V$ and $I$ bands.  The Red Giant Branch (RGB) of Fornax is located in the approximate magnitude range $18 < V < 21$ at a colour of $(V-I) \approx 1$.  To accurately map the Fornax RGB stars, errors less than 0.05 mag (an approximate signal-to-noise ratio of 25) were required down to magnitudes of $V=20$ and $I=19$.  Therefore, multiple images were recorded for each field with total exposure times of 2400s in $V$ and 1800s in $I$.  By coadding the images, effects such as cosmic rays and satellite trails were removed.  The exposure times for each field are listed in Table \ref{fornaxobservations}.  Poor seeing was compensated by taking an extra exposure of the relevant field (for example, see Field 15 in Table \ref{fornaxobservations}).  The tracking of the telescope was found to deteriorate with increasing zenith distance.  Therefore, if the desired field was located $\gtrsim 2$ hours from the meridian, the exposure times were halved and the number of exposures were doubled.  Additionally, during periods of good observing conditions (seeing approaching $1''$), guide stars were used to reduce the effect of telescope tracking on the star profiles.


Image reduction was accomplished using standard routines in the Image Reduction and Analysis Facility\footnote{IRAF is distributed by the National Optical Astronomy Observatories, which is operated by the Association of Universities for Research in Astronomy, Inc., under contract with the National Science Foundation.} program following the standard procedure.  The bias images and overscan region were subtracted from all object and flat field exposures.  All object images were divided by the appropriate combined twilight flat field, accounting for variations in CCD pixel sensitivity.  The images were also divided by a secondary dark sky flat, a median combination of long exposures from each of the sixteen pointings listed in Table \ref{fornaxobservations}.

Each image was split into its eight CCD component parts.  The eight CCD images were then individually registered with all other CCD images for that field, by comparing the pixel coordinates of several bright stars.  The CCD images were median combined to create the final object image, where the final images only included stars falling in the common areas of each exposure.  By ensuring that the maximum offset between each frame was less than 30 pixels, the amount of sky coverage lost was limited to $2\%$ of the WFI frame.  WFI has a `filling factor' of approximately $97.5\%$, therefore the Fornax survey covered approximately $95\%$ of the desired area.

An astrometric calibration was made using the first USNO CCD Astrograph Catalogue (UCAC1; \citealt{zach00}), which is estimated to have an average precision of 31 mas in the magnitude range $8 < R < 16$.  To test astrometric accuracy, stars were matched in the overlap regions of the Fornax survey.  This was accomplished by searching for a neighbour within $1.0''$ of each star, where the star and its neighbour were required to originate from different fields.  A total of 6556 matches were found in the overlapping $V$ images in the magnitude range $16 \le V \le 20$, while the algorithm yielded 3779 matches in the $I$ magnitude range $15 \le I \le 19$ ($V$ and $I$ are magnitudes on the standard system; see the following section).  Figs.\ \ref{coordsV} and \ref{coordsI} show the resulting difference in coordinates between these matched pairs.  The distributions are well fit by a Gaussian function (dotted lines), and are centered close to zero, indicating there are no systematic errors affecting the coordinates.  Also, the standard deviations of the best-fit Gaussian distributions give the astrometric error as $0.12''$ in RA and $0.10''$ in Dec.  These errors are not magnitude-dependent in the $V$ and $I$ ranges given above.

\section{Photometry}
Stellar magnitudes in all science images were measured using the DAOPHOT program \citep{stetson87} within IRAF, which uses an interactive PSF-fitting algorithm.  The DAOPHOT program searches for stars in the image using the PSF full-width half-maximum (FWHM) and standard deviation of the background (calculated as $\sigma_{bg} = \sqrt{N_{bg}}$, where $N_{bg}$ is the background level in electrons) as search parameters.  We searched each image for sources with a peak flux greater than $4\sigma_{bg}$ above the background level.  If the source was well-fit by an analytic Gaussian, the total flux within 25 pixels ($9.4''$) was measured to provide an aperture magnitude.  The twenty brightest stars in each image were examined by eye, and those with a `good looking' profile (that is, those stars without bright neighbours or nearby pixel defects) were median combined to produce an empirical PSF model for each CCD image.  The PSF model radius was 25 pixels.  The PSF was then fitted to the profile of each star in the coordinate list using a fitting radius of 6 pixels ($2.3''$) or $1-2$ half-width, half-maximum of the PSF, dependent on atmospheric seeing.  Selecting a fitting radius close to the PSF HWHM has a twofold effect: (i) the probability of interference from nearby stars is decreased, and; (ii) the effect of background noise on the stellar profile is reduced (especially in faint stars).  For images recorded during poor seeing, the DAOPHOT search algorithm was found to occasionally identify more than one centre to a star.  To remove such double detections, the magnitudes and centres of these `double stars' were averaged.

\subsection{Photometric Calibration}
The selection of candidate Fornax stars in the outer regions of the object was dependent on an accurate photometric calibration of the entire dataset.  For straightforward observations, a series of standard fields are imaged throughout the observing run, and the instrumental magnitudes of the standard stars are compared to their magnitudes on the standard system.  This results in a solution to the transformation equations, which are then used to shift the observations to the standard system.  However, the Fornax survey was a compilation of sixteen fields imaged in two filters throughout four observing runs, and the process of using several sets of standard star measurements to independently calibrate each field would be overly time consuming.  Also, the detection efficiency is not constant over the WFI array, hence an independent calibration of each CCD was required.

Therefore, the photometric calibration was divided into three parts.  First, an internal calibration of the WFI mosaic was conducted, using standard field obervations on each CCD during the December 2003 observing run.  This allowed a comparison of the zero-points of each CCD, shifting the photometry to the same instrumental photometric system.  Second, an inter-field calibration was achieved by comparing the instrumental magnitudes measured in the overlap regions between each field.  The zero-points of each field were adjusted to match the field F10, such that all fields were on the same photometric system.  Finally, a set of standard fields observed concurrent with F10 in October 2001 were analysed to solve the transformation equations.  Thus, the entire Fornax dataset was adjusted to the standard magnitude system.  A full description of these three procedures is contained below.  All standard fields were chosen from those provided by \citet{graham82} and \citet{landolt92}.  When possible, the standard magnitudes used throughout the following analysis came from the Canadian Astronomy Data Centre Standards web page\footnote{http://cadcwww.hia.nrc.ca/standards/} \citep{stetson00}.

\subsubsection{WFI Mosaic Calibration}
Each CCD  in the WFI array displays a significantly different gain (ranging approximately from 1.4 to 2.0 $e^-$/ADU), thus the zeropoint calibration in ADUs is CCD-dependent.  Consequently, the first step in the photometric calibration was to measure the photometric difference between the CCDs.  The E2 standard field \citep{graham82} was imaged on each CCD during the photometric night of 17th December 2003.  These observations was conducted over a small airmass range to maintain observing conditions for each CCD.  The images were recorded in sets of six, where each set consisted of three $V$ and three $I$ images with exposure times of 10, 20 and 30s.  Image reduction was achieved following the method described in the reduction section above.  Dark sky flat images were obtained by combining long-exposure Fornax images from the December 2003 observing run.

A measurement of a stellar magnitude can be accomplished using either a PSF-fitting technique or aperture photometry.  Although the PSF technique is effective for faint stars and/or a crowded stellar field, aperture photometry provides an accurate measure of the total light from a bright, uncrowded source such as a standard star.  Therefore, the instrumental magnitudes of all standard stars were measured using aperture photometry.  A series of concentric circles were placed on each standard star and the flux through each aperture was measured using the {\em phot} routine in IRAF.  The aperture magnitude ($v$ and $i$) of the standard star was then measured as the asymptote of these values.  A total of 336 aperture magitudes were measured for standard stars across the eight WFI CCDs from the December 2003 dataset.

These standard star magnitudes were used to provide a solution to the transformation equations,
\begin{equation} \label{transformationeqns}
\begin{array}{c}
V_{\mbox{\scriptsize st}}=v + Z_V - k_V X - c_V (v-i), \\
I_{\mbox{\scriptsize st}}=i + Z_I - k_I X - c_I (v-i),
\end{array}
\end{equation}
\noindent for each CCD.  The zeropoints ($Z_V$ and $Z_I$) and the colour terms ($c_V$ and $c_I$) are CCD-dependent, and are usually calculated using the measured aperture magnitudes ($v$ and $i$) and the known standard magnitudes ($V_{\mbox{\scriptsize st}}$ and $I_{\mbox{\scriptsize st}}$).  However, the E2 standard stars have a colour range of $0.6 < (V-I) < 1.0$, which is not large enough to provide an accurate value for the colour terms.  Hence, $c_V$ and $c_I$ were assumed to be zero, and the zeropoints ($Z_V$, $Z_I$) of each CCD were determined using the E2 standard star solutions.  An analysis of CCD 2 with a large standard star colour range (section \ref{photcal}) indicates colour terms of $\sim 0.01$ in $V$ and $I$, too small to significantly affect the photometry.  Hence, we assume colour dependence is negligible over the WFI array.

Figs.\ \ref{Vstandards_ccd} and \ref{Istandards_ccd} display the standard star observations for each CCD.  $\Delta V$ and $\Delta I$ are defined as the difference between instrumental (extinction-corrected) and standard magnitude,
\begin{equation} \label{deltamag}
\begin{array}{c}
\Delta V = v - k_VX - V_{\mbox{\scriptsize st}}, \\
\Delta I = i - k_IX - I_{\mbox{\scriptsize st}}.
\end{array}
\end{equation}
Therefore, under the assumption that the colour terms $c_V$ and $c_I$ are zero, the mean $\Delta V$ and $\Delta I$ values provide an independent measure of the zeropoints for each CCD.  These $Z_V$ and $Z_I$ values are listed in Table \ref{ccdcalib}, where the errors are the standard deviation of the mean of the points in Figs.\ \ref{Vstandards_ccd} and \ref{Istandards_ccd}.  Table \ref{ccdcalib} also includes values for $\Delta Z = Z_j-Z_2$, where $j$ is the CCD number.  These are calibrating values to convert all WFI magnitudes to a photometric system consistent with CCD 2.  For example, the $\Delta Z$ values in Table \ref{ccdcalib} indicate that instrumental stellar magnitudes measured in CCD 1 ($\Delta Z_I = -0.085 \pm 0.009$ mag) are significantly fainter compared to the remainder of the array in $I$ band.  This is the result of lower $I$ band quantum efficiency for this CCD.  All subsequent WFI photometry was adjusted by the appropriate $\Delta Z$ value, thus producing constant photometry across each WFI image.

\subsubsection{Inter-Field Calibration}

The selection of Fornax-candidate stars is dependent on a consistent photometric system across the entire mosiac of fields.  However, variable atmospheric conditions over several observing runs implies a different zeropoint for each field.  Also, the limited time at the telescope did not allow enough instrumental standard field observations to independently calibrate each field in the survey.  Therefore, a robust process was required to adjust the magnitudes in each field such that the survey constituted an internally calibrated dataset.  The previous section described the method to internally calibrate each WFI field such that magnitudes were not CCD-dependent.  This section outlines the process to calibrate all Fornax fields to the same photometric system.

\citet{liske03} describe the photometric calibration of the Millenium Galaxy Catalogue, a CCD-based imaging survey covering 37.5 deg${}^2$ along the celestial equator.  The internal calibration of the Fornax dataset followed the same process.  Stellar magnitudes for all Fornax fields were measured using the DAOPHOT PSF-fitting procedure, and those in the $10\%$ overlap region were matched based on astrometry to find multiply detected stars.  The overall zeropoint difference between fields was calculated as the average of these magnitude differences.  For example, Fig.\ \ref{field5calib} shows the magnitude difference measured between the fields F5 and F6 based on 120 and 133 doubly detected stars in $V$ and $I$ respectively.  The dashed lines represent the measured zeropoint difference between F5 and F6 in both $V$ and $I$ bands.  That is, the overall photometric shift between these fields was found to be,
\begin{displaymath}
\begin{array}{c}
V_5 - V_6 = 0.111 \pm 0.043, \\
I_5 - I_6 = -0.126 \pm 0.060, \\
\end{array}
\end{displaymath}
\noindent where the errors are the standard deviation of the distributions in Fig.\ \ref{field5calib}.  This process was repeated for all 24 overlap regions.  Note that these zeropoint differences were measured after adjusting all magnitudes by the values given in Table \ref{ccdcalib} to a CCD-independent photometric system.  The magnitude shift for each field was measured by averaging the zeropoint differences obtained for all adjacent fields.  Table \ref{fieldconv} lists the resulting values of $\Delta Z_V$ and $\Delta Z_I$, which were added to the zeropoint of each field to adjust them to the F10 photometric system.

Table \ref{fieldconv} does not include the errors for $\Delta Z_V$ and $\Delta Z_I$ due to the difficulty in estimating the propogation of errors across the mosaic of fields.  To estimate the internal calibration error the magnitude shifts listed in Table \ref{fieldconv} were applied, and the magnitude difference between stars matched in the overlap regions was calculated.  Ideally, $\Delta V$ and $\Delta I$ for all matched stellar pairs should be zero.  Fig.\ \ref{phothist} shows the distribution of $\Delta V$ and $\Delta I$ for all stars brighter than $V=20$ and $I=19$ .  The dotted lines are the best-fit Gaussian curves to the histograms, centred at $\Delta V = 0.005$ and $\Delta I = -0.001$ mag.  These two values are close to zero, which implies a constant photometric system across the Fornax dataset.  The standard deviation of these distributions give an internal photometric accuracy of $\sigma_V = 0.048$ and $\sigma_I = 0.051$ mag.  These errors include those contributed by the CCD zeropoint offset described in the previous section.

\subsubsection{Photometric Calibration} \label{photcal}
The previous two sections accounted for CCD-dependent and field-dependent photometry, resulting in a survey consistent with F10 to an accuracy of $0.05$ mag in $V$ and $I$.  F10 was observed on the 20th October 2001 concurrent with a large set of standard fields on CCD 2.  We describe below the analysis of these standard field images to calibrate the Fornax dataset with the standard system.  The instrumental magnitudes in $V$ and $I$ were measured using the aperture magnitude method described above.  $\Delta V$ and $\Delta I$ are the difference between instrumental (extinction-corrected) and standard magnitudes as defined in Equation \ref{deltamag}.


We examined $\Delta V$ and $\Delta I$ as a function of time to ensure no instrumental or atmospheric changes occurred throughout the night.  Additionally, these parameters displayed no significant dependence on airmass, hence the values of $k_V$ and $k_I$ provided by \citet{sung00} appear to accurately account for atmospheric extinction for the observations of this night.  Fig.\ \ref{20octstds} shows the standard star points from the night of October 20th.  Measuring a least-squares linear fit to the points resulted in the parameters,
\begin{displaymath}
\begin{array}{rl}
Z_V = 0.760 \pm 0.005, & c_V = -0.010 \pm 0.005 \\
Z_I = 1.297 \pm 0.006, & c_I = 0.015 \pm 0.006 \\
\end{array}
\end{displaymath}
\noindent with {\em rms} values of 0.022 and 0.026 respectively.  Therefore, including the DAOPHOT zeropoint of 23.5 gave the following solution to the transformation equations,
\begin{equation}
\label{transsol}
\begin{array}{c}
V=v + 22.740 - 0.16 X + 0.010 (v-i), \\
I=i + 22.203 - 0.086 X - 0.015 (v-i).
\end{array}
\end{equation}
The colour terms are close to zero in both $V$ and $I$.  Indeed, a zero slope line in Fig.\ \ref{20octstds} does not appear to adversely afffect the fit, giving almost identical {\em rms} values in $V$ and $I$.  That is, CCD 2 does not display a strong colour dependence.  Hence, our observations indicate that the WFI instrument gives photometry consistent with the Cousins $V$ and $I$ system.  Within the limited colour range, the data presented in Figs.\ \ref{Vstandards_ccd} and \ref{Istandards_ccd} support this assumption.  Hence, we use the transformation solution above (Equation \ref{transsol}) to shift the Fornax dataset to the standard system.

\subsection{Completeness}
Stars in the overlap regions were combined by searching for a neighbouring star within $0.5''$ for all stars in both the $V$ and $I$ datasets.  Based on the astrometry errors given above ($0.10''$ in $V$ and $0.12''$ in $I$), we expected this search radius to yield a match for $>99\%$ of all multiply imaged stars.  We removed the multiple detections by averaging the magnitudes measured for such stars.  Hence, this resulted in two maps of the Fornax region in $V$ and $I$ each covering an area approximately $3.1^{\circ} \times 3.1^{\circ}$ on the sky, extending approximately $0.5r_t$ beyond the tidal radius of Fornax \citep{m98}.  To join these two maps into a single dataset, stars in $V$ and $I$ were matched based on astrometry using a search radius of $0.5''$.  Fig.\ \ref{fornaxcmd} (left) shows the CMD for the inner $30'$ of Fornax.  The errorbars are the {\em rms} of the DAOPHOT errors shown in Fig.\ \ref{fornaxcmd} (right).  The RGB lies in the approximate range $18 < V < 21$, and the upper 0.5 mag of the red clump can be seen below $V \sim 21$.  The four inner fields which contributed to this figure were all imaged in relatively good conditions (see Table \ref{fornaxobservations}) hence Fig.\ \ref{fornaxcmd} represents the best depth achieved for the Fornax dataset.

Using the CMD to select candidate Fornax stars is dependent on the photometric completeness of the survey.  Each field was observed with the intention of reasonable photometry (errors $< 0.05$ mag) down to $V=20$ and $I=19$.  However this was dependent on observing conditions.  Therefore, a luminosity function was constructed for each field in $V$ and $I$.  The photometric limit of each field was set 0.2 mag brighter than the turnover magnitude of the luminosity function.  To ensure this method provided an accurate measure of the completeness limit for each field, we simulated a stellar population resembling the Fornax RGB, and populated each field with $\sim$5000 artificial stars.  The photometric routines described above were repeated over the dataset, and it was found that at least $94\%$ of the artificial stars were recovered in all fields down to their completeness limit.  In the $I$-band, all fields were complete to $I=19$.  However, in the $V$-band, the field F13 had the brightest limit of $V=19.4$, while the photometry for the remaining fields was complete above $V=20$.  Thus, we repeated the artificial star experiment for F13 down to $V=20$, and found it to be $89 \pm 2\%$ complete.  Hence, a photometric limit of $V=20$ was imposed on the entire dataset, and these are marked as the dashed line in Fig.\ \ref{fornaxcmdlines}.  We discuss possible incompleteness effects in section \ref{denfn} below.

\section{Results} \label{results}

Fig.\ \ref{fornaxcmdlines} shows the CMD region used to select candidate Fornax stars down to the nominal photometric limit of $V=20$.  The selection region was extended approximately 0.15 magnitudes in colour to either side of the RGB, to account for photometric errors ($\sim 0.05$ mag at $V \sim 19.5$ in F13, which had the brightest limiting magnitude) and errors resulting from the inter-field calibration (measured as $0.05$ mag in $V$ and $I$).  Note that the zero-point calibration is irrelevant to this process.  Selecting stars based on colour and magnitude increases the ratio of Fornax stars to background/foreground sources, thereby improving the signal-to-noise ratio of Fornax stars in the outer regions.

Fig.\ \ref{fornaxrgbxy} displays the spatial distribution of these CMD-selected stars, where $\Delta \alpha$ and $\Delta \delta$ are calculated relative to the central coordinates of Fornax and the inner and outer lines are the core and nominal tidal ellipses of Fornax respectively \citep{m98}.  The centre of these ellipses come from the analysis in the next section.  There are several important points to note about this figure.  Firstly, two fields to the lower left appear shifted by approximately $0.15^{\circ}$ towards the centre of the mosaic, causing a slight gap in the survey region.  These are the fields F9 and F14, which were part of the initial observations recorded in October 2001 (see Table \ref{fornaxobservations}).  An analysis of these fields revealed an excessive overlap region; the initial overlap fraction for the mosaic was $15\%$.  This value was decreased to $10\%$ for the ensuing observations, increasing the spatial coverage of the survey without significantly altering the inter-field photometric calibration.  Re-observation of F9 and F14 with the altered field placement was not possible in the remaining observing time.

There are also some possible systematic effects present in F13, located to the lower left of Fig.\ \ref{fornaxrgbxy}.  Although artificial star tests indicate the photometry in this field is $\sim$$90\%$ complete down to $V=20$, it is difficult to measure whether the variation in stellar number density is real structure or systematic effects.  Hence, we include F13 in the following analysis under the assumption that any possible systematic or completion effects are negligible.  A full discussion of F13 is included in the next section.

Perhaps the most apparent feature in Fig.\ \ref{fornaxrgbxy} is a region of increased stellar density located approximately $1.3^{\circ}$ north-west from the centre of Fornax.  This apparent overdensity is situated in F4 {\em outside} the nominal tidal radius.   An inspection of the CCD images for F4 also indicates all the stars within this feature are real stellar sources, while the star finding algorithm DAOPHOT does not appear to have missed any stars in the remainder of the field.  Furthermore, given the high Galactic latitude of Fornax, we do not expect this feature to be caused by dust extinction.  The IRAS $100 \mu$m \citep{schlegel98} map in Fig.\ \ref{fornax_dust} does not show any significant dust structure in the region of this feature.

A full discussion of this feature is presented in the next section, however here we attempt to measure its clarity compared to the background population.  The luminosity functions indicated six fields, including F4, were complete to $V=20.7$ at the Fornax RGB, represented by the lower dashed line in Fig.\ \ref{fornaxcmdlines}.  Utilising the deeper photometry of F4, we refined the CMD selection region (shown in Fig.\ \ref{fornaxcmdlines}) to give the greatest density contrast of this feature compared to the remainder of the field F4.  This was achieved by adjusting the width of the selection region in increments of 0.01 mag, selecting those stars which satisfied this CMD constraint, and then measuring the density of stars in the overdense feature ($\rho_{\mbox{\scriptsize feature}}$) and the remainder of F4 ($\rho_{\mbox{\scriptsize F4}}$).  Thus, the CMD selection range which maximised the `signal' from the overdense region was that with the highest ratio, $\rho_{\mbox{\scriptsize feature}}/\rho_{\mbox{\scriptsize F4}}$.  We found that compressing both sides of the selection region in Fig.\ \ref{fornaxcmdlines} by 0.16 magnitudes resulted in a maximum density contrast, where $\rho_{\mbox{\scriptsize feature}} = 0.46 \pm 0.04$, and $\rho_{\mbox{\scriptsize F4}} = 0.16 \pm 0.01$ stars/arcmin${}^2$ (uncertainties represent Poisson noise).  That is, the overdense region contains a stellar density $2.9 \pm 0.3$ times that of the field population.  The probability that this enhanced region represents a random fluctuation in the field population is minimal, and hence it warrants further investigation.  This refined CMD selection region is shown in Fig.\ \ref{clumpcmd}, and the spatial distribution of the selected stars is shown in Fig.\ \ref{fornaxrgbdeepxy} for the six complete fields only.  The feature has a greater clarity than in Fig.\ \ref{fornaxrgbxy} and may contain two main components.  This structure is further discussed in the next section.

One test for extra-tidal structure is provided by the stellar radial profile.  \citet{king62,king66} models have been found to provide a satisfactory description of the structure of nearby dSph galaxies (for example, \citealt{ih95}).  These models assume the system can be be represented as a truncated isothermal sphere.  That is, the stars have equal mass with a Maxwellian velocity distribution, and the system is in a relaxed state with a limiting radius determined by the tidal field of the host galaxy.  Also, single-component King models assume the distribution of mass is matched by the distribution of light.  Despite these simplifying assumptions, King models are a useful tool for the analysis of the local dSph population.  If the star counts in the outer region exceed the best-fitting King profile, this is a possible indication that the structure of the system has been inflated by interacting with the host galaxy.  For example, the Ursa Minor dSph possesses a significant population of extra-tidal stars \citep{mart-del01,palma03} and displays signs of structural distortion \citep{ih95,kleyna98} possibly due to the Galactic tidal field.

\citet{ih95} derived a tidal radius of $71 \pm 4$ arcmin for Fornax, with some indication of extra-tidal structure.  However, the CCD survey by \citet{walcher03} found a tidal radius of $98'$, significantly larger than previous estimates.  Further, Walcher et al.\ found no excess star counts at large radii, though they note that the survey did not cover a large enough area to make an accurate measurement.  We created a radial profile for the red giant stars shown in Fig.\ \ref{fornaxrgbxy} using the Fornax centre, ellipticity and position angle listed by \citet{m98}.  The profile is shown in Fig.\ \ref{radial}, where the dashed line represents the King model from a least-squares fit.  The three King parameters, core radius ($r_c$), tidal radius ($r_t$) and central density ($k$), were derived using a simple least-squares fit in logarithmic space to the observed data points.

We have measured the core and tidal radii to be $r_c = 16.0 \pm 1.5$ and $r_t = 64.0 \pm 2.0$ arcmin respectively.  These values fall within the range defined by previous measurements (see \citealt{ih95} for a review).  The star counts at large radii ($r > 50'$) exceed the King model by amounts significantly larger than the uncertainties, providing strong evidence for extra-tidal structure.  Although \citet{ih95} found evidence for a break in the profile, it occurred at a larger radius in their data.  This effect may be due to increased contamination from field stars in the Irwin \& Hatzidimitriou survey compared to the current study.  Furthermore, the survey of \citet{walcher03} did not cover a large region outside the Fornax system, which may have resulted in an inaccurate estimate of the field star density; consequently, Walcher et al.\ may have interpreted the excess of stars at large radii as an indication of a large tidal radius.

Although we were able to remove a significant number of the field stars, \citet{ih95} possessed photometry at a greater depth than the current survey.  Therefore, unless stated otherwise, we use the Irwin \& Hatzidimitriou tidal radius as the nominal value for the remainder of this paper.  We note that the choice of tidal radius has no substantial effect on our final results.  To investigate the possibility of Fornax stars beyond the nominal tidal radius, we conducted two independent statistical analyses of the extra-tidal region to search for stellar structure, the results of which are described in the next section.

\section{Outer Structure} \label{outerstr}

\subsection{Density Function} \label{denfn}
We created a density probability function of the region outside the nominal Fornax tidal radius using a Monte Carlo evaluation.  This technique consisted of placing $10^6$ circles (radius $12'$) at random coordinates in the $V \le 20$ dataset (Fig.\ \ref{fornaxrgbxy}) and counting the number of stars in each circle.  Only circles whose centre was located outside the nominal Fornax tidal radius and within the dataset boundary contributed to this process.  The resulting histogram (Fig.\ \ref{mccontour}) details the probability of measuring a given stellar density.  In effect, by completing a large number of iterations, a binomial function describing the density of stars outside the Fornax tidal radius was obtained.  The probability density function was sensitive to structure of minimum size $\sim$$10'$.  We have tested this method on artificial datasets (Coleman et al.\ in prep), and have found it to be sensitive to large-scale structures with a mean density greater than $1.5\sigma_{\mbox{\scriptsize field}}$ (this parameter is described below) above that of the mean field stellar density.

A uniform distribution of randomly placed points is expected to yield a Gaussian function.  However, Fig.\ \ref{mccontour} indicates the probability function around Fornax is more complex than a simple random distribution.  The histogram (solid black line) appears dominated by two Gaussian populations centred at $\rho=350$ and $525$ stars/deg${}^2$.  To determine the number of field stars expected in this region of the sky at the Fornax RGB colour-magnitude range, we consulted the \citet{rat85} estimation, based on the \citet{bahcall80} Galactic model.  \citet{rat85} list the expected field star density for a colour range of $0.8 < (B-V) < 1.3$ and magnitude range of $17 < V < 21$ towards the Local Group dwarf galaxies.  Although this does not exactly correspond to our colour-magnitude range, we used our Sculptor photometric dataset (Coleman et al. in prep) to normalise the field star density for the Fornax RGB selection range.  Hence, the \citet{bahcall80} Galaxy model indicates a field star density of 368 stars/deg${}^2$ for our desired colour-magnitude selection towards Fornax, which is very close to the first peak in the density distribution.  \citet{rat85} estimate a star count error of $25\%$, hence we conclude the low-density peak in Fig.\ \ref{mccontour} corresponds to the true field star density.

The best-fit Gaussian function to the field star population in Fig.\ \ref{mccontour} is centred at $\rho=350$ stars/deg${}^2$ and has a standard deviation of $\sigma_{\mbox{\scriptsize field}}=38.2$ stars/deg${}^2$.  Correspondingly, the second peak (representing the medium-density population) centred approximately at $\rho=525$ stars/deg${}^2$ has a standard deviation of $\sigma_{\mbox{\scriptsize med}}=169$ stars/deg${}^2$.  This second peak is separated by $4.6\sigma_{\mbox{\scriptsize field}}$ from the field population.  Note that the standard deviation of the second peak is significantly larger than we would expect for a uniform distribution of stars, indicating this medium-density population contains a wide range of stellar densities, possibly indicating structure.  The red line in Fig.\ \ref{mccontour} is the sum of the two Gaussian distributions.  The greyscale bar defines the density contours in Fig.\ \ref{fornaxcontour_outer}, which is a filled contour diagram of Fornax RGB-selected stars (Fig.\ \ref{fornaxrgbxy}) outside the nominal tidal radius.

Figs. \ref{mccontour} and \ref{fornaxcontour_outer} indicate that the regions North-East and South-West of Fornax contain {\em only} the field star population ($\rho=350$ stars/deg${}^2$, approximately the first density level).  We conclude that there is no significant population of Fornax RGB stars in these regions.  However, the medium-density population ($\rho=525$ stars/deg${}^2$) is confined to the regions to the North-West and South-East of Fornax, which are located on the minor axis of Fornax.  From the density and standard deviation of the field star population, all regions with a density greater than $450$ stars/deg${}^2$ (the second density level) are distinct from the field star population by at least $2.7\sigma_{\mbox{\scriptsize field}}$.  Hence, we find there is a significant overdensity of Fornax RGB stars in the extra-tidal regions of the NW/SE quadrants.

A third population is visible in Fig.\ \ref{mccontour} with a peak density at $1150$ stars/deg${}^2$, more than twice that of the medium-density population.  This density peak corresponds to the relatively high-density structure located approximately $1.3^{\circ}$ NW from the centre of Fornax (Fig.\ \ref{fornaxcontour_outer}).  We find it is a $3.6 \sigma_{\mbox{\scriptsize med}}$ increase on medium-density population.  However, if we assume the true background is the field star population as determined above, then this region represents a $220\%$ increase on the background, or approximately a $21\sigma_{\mbox{\scriptsize field}}$ result.   Therefore, we have detected an excess of Fornax RGB stars in the NW/SE regions outside the tidal radius, plus a {\em further} excess of stars $1.3^{\circ}$ NW from the centre of Fornax.  With no corresponding excess in the NE/SW regions, we conclude that Fornax extra-tidal stars are located preferentially along the minor axis.

To further test this result, we divided the extra-tidal region into two datasets.  The first consisted of the field star-dominated quadrants to the NE and SW, and the second dataset comprised the overdense regions to the NW and SE of Fornax.  The probability density function was evaluated for these two subsets using the same process described above, and the results are illustrated in Fig.\ \ref{mcquads}.  The upper panel demonstrates that the NE/SW quadrants are dominated by a single population density which corresponds to the \citet{rat85} field star counts.  This function displays a tail extending to higher densities, which represents a small amount of `leakage' from the NW/SE quadrants.  The lower panel describes the density of the overdense NW/SE quadrants.  The peak from the NW overdensity (indicated by the arrow) is significantly increased compared to that shown in Fig.\ \ref{mccontour}.  This substantiates our inference that the extra-tidal region of Fornax contains an increased density of stars along the minor axis.

\subsubsection{Completeness Effects}

The field F13 is located towards the bottom left of the contour plot, and the majority of this field has a Fornax RGB-selected stellar density of $\sim$$350$ stars/deg${}^2$.  However, we estimated above that F13 is $89\% \pm 2\%$ complete down to $V=20$.  Therefore, we calculated the density of Fornax RGB-selected stars in F13 to be $\sim$$400$ stars/deg${}^2$.  Although this is approximately $30\%$ less than the density of stars in the two neighbouring fields (Fig.\ \ref{fornaxcontour_outer}) it is a $1.4 \sigma_{\mbox{\scriptsize field}}$ increase on the field star population, and does not contradict our result of an excess Fornax RGB population in the SE quadrant of the survey.

Fig.\ \ref{fornax_dust} shows the distribution of the reddening parameter $E(B-V)$ over the Fornax field as determined by \citet{schlegel98}.  Although the majority of the survey region has a reddening of $E(B-V) \sim 0.02$ mag, the dark region towards the lower right corresponds to $E(B-V) \sim 0.07$ mag.  Thus, the visual extinction difference between this SW region compared to the remainder of the survey is approximately 0.2 magnitudes.  To determine if this caused significant incompleteness effects, we adjusted the magnitudes and colours measured in the SW quadrant using these reddening and extinction differences, and then re-applied the CMD selection range defined in Fig.\ \ref{fornaxcmdlines}.  The SW extra-tidal star counts were found to increase by approximately $5\%$, which is negligible within the uncertainties.  Also, this increase does not bring the SW stellar density within the range of the NW/SE density.  Hence, differential extinction across the Fornax field does not substantially affect our result.

\subsection{Angular Correlation Function}
A common method of determining galaxy clustering is to measure the two-point correlation function, which is equivalent to the Fourier transform of the power spectrum of a distribution of point sources \citep{peebles80}.  A sample of galaxies with Poisson spatial distribution produces a flat angular correlation function of value approximately zero when plotted against angular scale.  However, if the distribution of galaxies contains structure on some scale, the angular correlation function will have a non-zero value at that scale.  We measured the angular correlation function in the Fornax RGB dataset to further quantify the level of substructure outside the tidal radius.  A description of the angular correlation function and its sensitivity to stellar clustering is given in the companion paper for the Scl dSph galaxy (Coleman et al.\ in prep.).

In the previous section, our results indicated a duality in the distribution of RGB-selected stars beyond the Fornax tidal radius: the NE/SW quadrants contained only the field star population, while the NW/SE quadrants displayed an excess of RGB stars.  To further test this result, we measured the angular correlation function individually for these four quadrants in logarithmic bins of width $\Delta \log \theta = 0.05$, where $\theta$ is in arcmin.  The function was measured over an angular scale of $40'$, or approximately half the width of each quadrant.  Stars inside the nominal tidal radius were excluded from this analysis.

The {\em observed} angular correlation functions for the NE/SW quadrants are shown in Fig.\ \ref{corrfn_boring}, where the dashed line represents $w_{\mbox{\scriptsize obs}}(\theta) = 0$.  These is some scatter about this line, however it provides an adequate overall description of the data.  Hence, we conclude that these regions have a flat angular correlation function and do not contain any sigificant structure.  These quadrants are situated on the major axis of Fornax and we do not have the same relative spatial coverage as compared to those on the minor axis.  Consequently, they contain less extra-tidal stars (430 in the NE and 502 in the SW quadrant) than the NW/SE sections (1274 and 1087 stars respectively), and therefore the correlation function uncertainties are larger.  However, there is little evidence for a non-zero $w(\theta)$ in the NE/SW quadrants.

In contrast, the angular correlation functions for the NW/SE sections (Fig.\ \ref{corrfn_interesting}) were both well-fit by a power law.  The $w(\theta)$ points measured from the NW quadrant follow a $\theta^{-0.36}$ power law, while that for the SE section is of the form $\theta^{-0.26}$.  The relatively large number of extra-tidal stars in these sections (and thus the small error bars) suggests both these relations are well constrained, hence these two quadrants contain a significant level of stellar clustering.  Note that the function in Fig.\ \ref{corrfn_interesting}(a) has an amplitude $A_w = 0.88$ compared to 0.53 in (b), indicating the stars in the NW section are more clustered than those in the SE.  This is also reflected in the power law slopes ($\beta = 0.36$ in the NW as compared to $0.26$ in the SE).  Furthermore, at angular scales $> 20'$, the data points in Fig.\ \ref{corrfn_interesting}(a) exceed the best-fit power law significantly beyond the Poisson errors.  This is representative of the high-density structure located $1.3^{\circ}$ NW from the centre of Fornax.  As a result, while the distribution of stars in the SE quadrant is significantly structured, the stellar clustering is more pronounced in the NW.  Therefore, we conclude that the Fornax dSph contains extra-tidal structure along the minor axis, with no corresponding structure in the major axis quadrants.  This corroborates the result described in the previous section.

\subsection{Dual Radial Profiles}
To further test the duality of the extra-tidal region of Fornax, we split the dataset into two regions; the NW/SE quadrants and the NE/SW quadrants.  The analyses above found that the second region contained a uniform-density extra-tidal stellar population (field stars), while the first region contained an increased density of stars beyond the nominal tidal radius.  A radial profile was created for each region using the same method as described in \S \ref{results}, with the results shown in Fig.\ \ref{radialsplit}.  The left-hand panel contains the data for the overdense NW/SE regions, providing a clear sign of the presence of extra-tidal stars beyond $r \sim 50'$.  In contrast, the datapoints for the second region (right panel) do not deviate from the best-fitting King model.  This supports the results described above.  Both radial profiles displayed a core ($r_c \sim 15'$) and tidal radius ($r_t \sim 66'$) essentially equivalent to those derived in \S \ref{results}.  These results, combined with those of the density function and angular correlation function above present the clearest evidence to date of extra-tidal structure in Fornax.

\subsection{Properties of the NW Shell} \label{nwshell}
From these analyses we found that Fornax displays significant extra-tidal structure.  We now focus on the shell-like feature located in the F4 field to the NW of the system, using the data down to $V=20.7$.  Fig.\ \ref{clumpcmd} shows the colour-magnitude data for all stars in this clump, demonstrating that there is a significant population of stars overlaying the Fornax RGB.  

We have examined why this feature has evaded detection until now.  There have been three wide-field studies of Fornax previous to the current work.  The surveys by \citet{eskridge88a,eskridge88b} and \citet{walcher03} did not include the overdense region, though \citet{ih95} covered a $5.1 \times 5.1$ deg${}^2$ area centred on Fornax, and therefore have data for this feature.  The main difference between that survey and the current one is the colour-magnitude star selection technique, which significantly decreases contamination from the field population; \citet{ih95} had access to a single `blue' (IIIaJ) photographic plate digitised through the APM facility, and were therefore unable to make a colour selection.  Although the feature is visible in their contour plot (Fig.\ 1 in \citealt{ih95}), it does not display the same clarity as seen in Fig.\ \ref{fornaxrgbdeepxy}.  For comparison, we retrieved the APM data\footnote{http://www.ast.cam.ac.uk/$\%$7Eapmcat/} for the Fornax region, and conducted the same density analysis as detailed above.  We find that the overdense region contains a stellar density approximately $1.3 \pm 0.1$ times that of the field population, compared to value of $2.9 \pm 0.3$ derived in the current survey (see \S \ref{results}).  Indeed, we measured the standard deviation of the APM field population around Fornax, which yielded the significance measurement of the shell feature to be $\sim$$1.5\sigma_{\mbox{\scriptsize APM}}$, as compared to $3.6\sigma_{\mbox{\scriptsize med}}$ from the current survey.  Also, our analysis of the APM data found little evidence of the lobed structures to the NW and SE of Fornax we described above.  Consequently, the colour-magnitude selection technique has helped distinguish these extra-tidal features.

\begin{figure}
\centerline{\hbox{
\includegraphics[height=190pt]{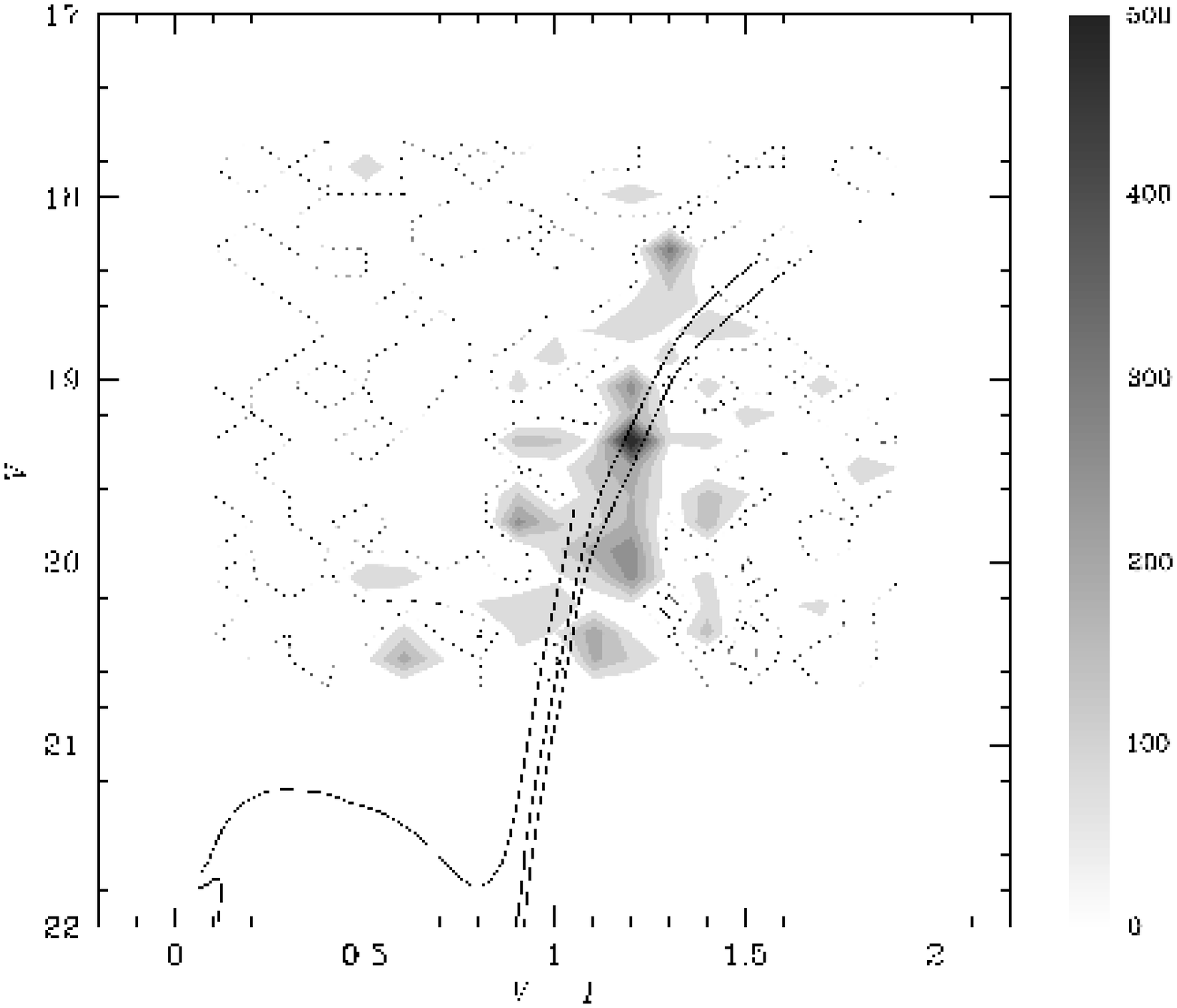}
\hspace{0.5cm}
\includegraphics[height=190pt]{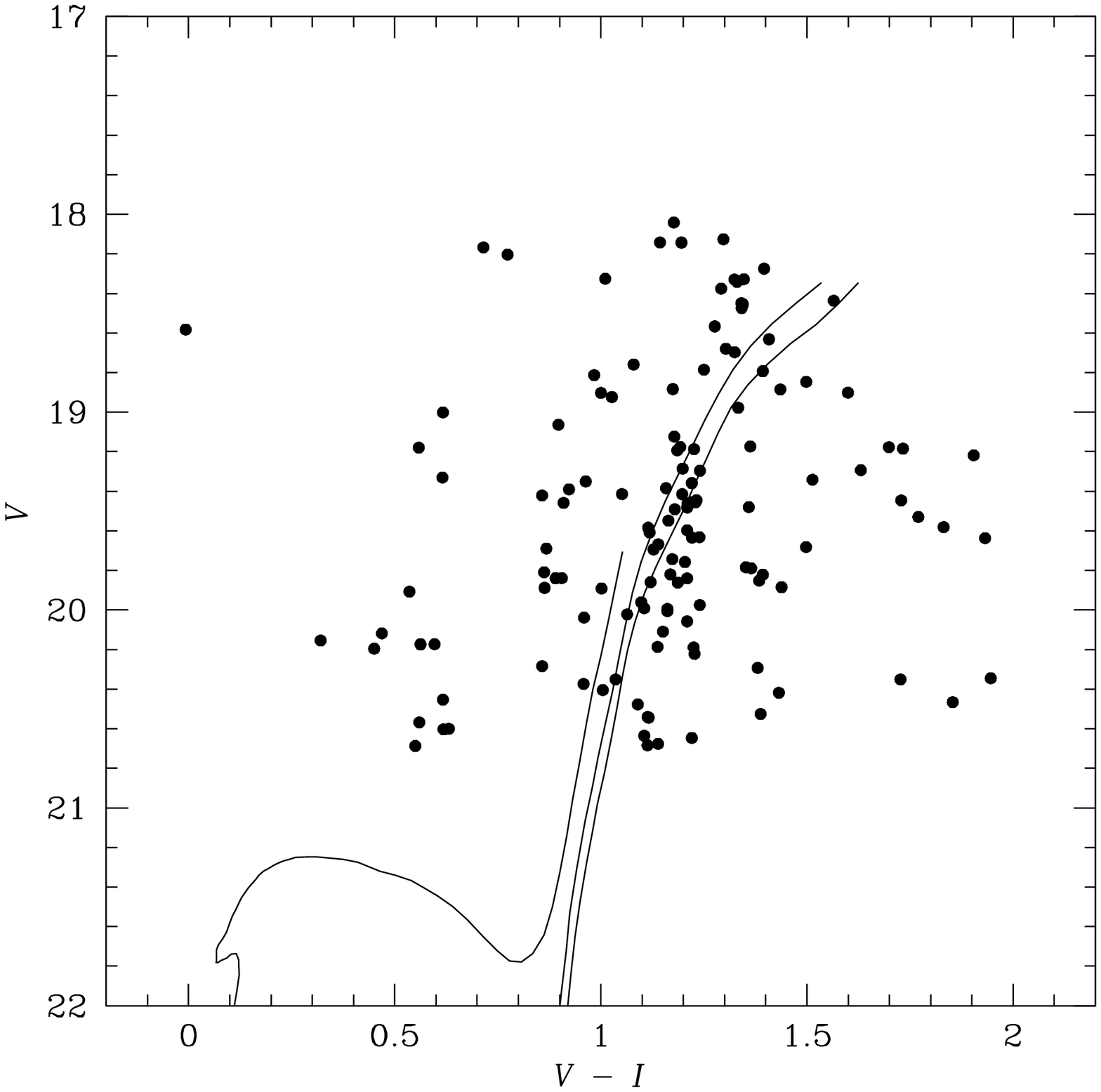}
}}
\figcaption[Field-subtracted CMD for the NW shell]{{\em Left panel:}  The background-subtracted CMD density function for the overdense region, $\Phi_{\mbox{\scriptsize feature}}$.  The faint limit of the function occurs at $V=20.7$.  Overplotted are Yonsei-Yale isochrones for ages of 1, 2 and 3 Gyr, and an assumed metallicity [Fe/H] = $-1.0$.  The density column at the right hand side defines the stellar density in the CMD in stars/mag${}^2$.  The dotted contour represents a stellar density of zero, and the dashed contour traces a density of approximately $-60$ stars/mag${}^2$. {\em Right panel:} A typical background-subtracted CMD for the overdense region.  This is an interpolation of $\Phi_{\mbox{\scriptsize feature}}$ after placing the appropriate number of stars in each cell. \label{cmdregsub}}
\end{figure}

To investigate the stellar population belonging to the NW overdensity, we have subtracted the field colour-magnitude data from that of overdensity following the method described in \citet{coleman04a}.  In summary, we created a CMD density function for the field population ($\Phi_{\mbox{\scriptsize field}}$) by dividing the colour-magnitude plane into a grid of $20 \times 20$ cells and counting the number of stars in each cell.  The process was repeated for the overdense region, resulting in the density function $\Phi_{\mbox{\scriptsize region}}$.  These were normalised, and the CMD density function of the feature was calculated as the difference between the two,
\begin{displaymath}
\Phi_{\mbox{\scriptsize feature}} = \Phi_{\mbox{\scriptsize region}} - \Phi_{\mbox{\scriptsize field}}.
\end{displaymath}
\noindent A greyscale representation of this function is shown in the left panel of Fig.\ \ref{cmdregsub}.  To aid in the visualisation of this field-subtracted density function, we ran several Monte Carlo simulations placing the appropriate number of stars in each cell with a random colour and magnitude within the cell boundaries.  A typical result is shown in the right panel of Fig.\ \ref{cmdregsub}.  The overlaid isochrones have been calculated using the Yonsei-Yale algorithm \citep{yi01,kim02}, and represent a stellar population ([Fe/H] $= -1.0$) with an age of 1, 2 and 3 Gyr respectively.  There is a density of stars near these isochrones, and they do not preclude the 2 Gyr age assigned to the inner shell based on young main sequence stars \citep{coleman04a}.  Deeper photometry is required to further investigate the stellar population in this feature.

It is worth noting that the metallicity of [Fe/H] $= -1.0$ was obtained from isochrones fitted to the original structure located near the Fornax core \citep{coleman04a}.  Hence, this value contains considerable uncertainty.  We have used the same isochrones for the NW feature to remain consistent with the analysis of the original shell, however a full spectroscopic analysis is required to accurately measure this parameter.  If these structures arose out of the collision between Fornax and a smaller companion, and given the correlation displayed by Local Group dwarf galaxies between size and luminosity (Fig.\ 7 in \citealt{m98}), it would be reasonable to expect these features to display a mean metallicity less than that of Fornax.

We found that the integrated light of all stars in the field-subtracted CMD gave a total magnitude of $m_V = 14.0 \pm 0.1$.  That is, assuming the feature lies at the same distance modulus as Fornax [$(m-M)=20.70 \pm 0.12$; \citealt{m98}] it has an absolute magnitude of $M_V=-6.7 \pm 0.2$.  For comparison, this lies within the luminosity range defined by the five Fornax globular clusters \citep{mackey03}.  However, if we assume the initial mass function of this feature follows a Salpeter law, then the stars below our survey limit will account for $\sim$$30\%$ of the total luminosity in this feature (depending on the age of the stellar population).  This light is not included in the integrated magnitude above, hence $M_V=-6.7 \pm 0.2$ represents an upper limit.  If we have missed approximately $30\%$ of the light, then we estimate the absolute integrated magnitude of the NW feature to be $M_V \sim -7$.  This is substantially more luminous than the original Fornax shell located just outside the core radius, which possessed an integrated absolute magnitude of $M_V \approx -4.2 \pm 0.6$.

\section{Inner Structure}
The earliest large-scale survey of Fornax by \citet{hodge61b} revealed the internal structure of this dSph to be quite complex, with an ellipticity ranging from $\sim$0.2 in the central region to $\sim$0.35 for the outer contours.  This result has since been confirmed by \citet{eskridge88a,eskridge88b}.  Perhaps the most comprehensive structural study of Fornax was made by \citet{ih95} using a `blue' photographic plate.  By subtracting an idealised elliptical profile from that of Fornax, they were able to confirm the asymmetry in the inner profile discovered by \citet{hodge61b}.  That is, the density contours were more closely spaced on the eastern side of the major axis than in the west.

The inner structure of a satellite system can be distorted by the gravitational potential of the host galaxy.  Simulations of a small stellar system orbiting a larger body (for example, \citealt{helmi01,mayer01,johnston02}) have found that the structure of the satellite often displays isophotal twisting, such that the ellipticity increases towards the outer regions, and is often accompanied by a changing position angle.  These effects have been observed in the Ursa Minor dSph \citep{mart-del01,palma03} which appears to be experiencing significant perturbation from the Galaxy.  Consquently, we examined the inner structure of Fornax for similar effects.


To reproduce a contour map of Fornax, we convolved each Fornax RGB star with a Gaussian using a radius of $40''$.  Fig.\ \ref{fornaxcontour} displays the resulting distribution of light overlaid with contours of smoothing length $3.0'$.  The contours are logarithmically spaced, where the first contour represents a stellar density $3.3\sigma_{\mbox{\scriptsize field}}$ above the field star population.  Note the overdensity of stars in the NW/SE sections (see \S \ref{outerstr}), including the feature located approximately $1.3^{\circ}$ NW from the centre of Fornax.  Further, the dependence of ellipticity on radius noted in previous analyses can be seen, and the position angle of these ellipses appears to change with radius.  A full derivation of these effects is given below.

\subsection{Radial Dependence of Structure}
We measured the radial dependence of the structure of Fornax using the output from the Monte Carlo algorithm described in \S \ref{denfn}.  This algorithm effectively smoothed the distribution of stars displayed in Fig.\ \ref{fornaxrgbxy} (which were selected using the Fornax RGB down to $V=20$).  Ellipses were placed at major axis radii $5, 6, \dots , 60$ arcmin.  At each radius, we iteratively measured the best-fitting centre, ellipticity and position angle.  Fornax has a nominal tidal radius of $\sim 65 - 70'$, though beyond $r=60'$ the number density of stars was too small to accurately fit an ellipse.  The results were then placed in $5'$ radial bins and are shown in Fig.\ \ref{lscontour}.  The dashed lines indicate the central coordinates, ellipticity and position angle of Fornax from \citet{m98}, and the errors are the standard deviation of each bin.

The upper two plots in Fig.\ \ref{lscontour} show the shift in central coordinates of Fornax with major axis radius.  Fig.\ \ref{lscontour}(b) indicates that despite the large difference between our central declination coordinates and those listed for Fornax ($\Delta \delta \approx -4'$) there appears to be little dependence of this value on radius.  In contrast, the right ascension (Fig.\ \ref{lscontour}(a)) demonstrates significant radial dependence, such that the centre of Fornax moves by $3'$ west within the inner $20'$ of Fornax.  This confirms the result of \citet{ih95} that the eastern contours are more closely spaced than those on the west.  However, we notice this effect reverses at larger radii, when the RA centre shifts back towards the east.  Stetson et al.\ (1998) noted that the centre appeared to move South by approximately $1.5'$ within the $15'$ from the centre of Fornax.  We find little evidence for this trend, though it is worth noting that Stetson et al.\ were able to access a significantly deeper dataset consisting of all stellar types, whereas we have only considered the distribution of RGB stars.  Thus, the results of Stetson et al.\ are most likely influenced by the young MS population forming a significant portion of the stars towards the centre of Fornax.  If we take the values at $r = 20'$ in Fig.\ \ref{lscontour}, then the centre of Fornax is $\alpha=02^h 39^m 47.9^s$, $\delta=-34^{\circ} 31' 01''$ (J2000.0), which is within $\sim$$2'$ of previous values \citep{hodge74,demers87,ih95,stetson98}.

The position angle is measured from North to East.  Fig.\ \ref{lscontour}(c) indicates that our result is consistent with \citet{m98}, who finds the position angle of the major axis of Fornax to be $48 \pm 6^{\circ}$.  The position angle of Fornax may be decreasing with radius.  Inspection of the contour plot in Fig.\ \ref{fornaxcontour}, supports this interpretation.  \citet{ih95} noted the ellipticity of Fornax appeared to increase with increasing radius, and Fig.\ \ref{lscontour}(d) does reflect this trend for the inner $35'$ of Fornax.  However, beyond this radius the ellipticity decreases with radius: the middle radii of Fornax ($15' \lesssim r \lesssim 40'$) are quite elliptical ($e \approx 0.25$), while inner and outer regions are more circular.  \citet{m98} cites the ellipticity of Fornax as $0.31 \pm 0.03$, which is consistent with our result at mid-range radii.

\section{Discussion}
The overdensity described in \S\S \ \ref{results} and \ref{outerstr} bears a striking resemblance to the concentric shells previously discovered in large galaxies \citep{malin80,malin83,schweizer80}.  Numerical modelling by \citet{hernquist88,hernquist89} demonstrated that shell structure is a common remnant of the slow collision between a large galaxy and a lower mass companion.  After the interaction, material from the companion galaxy is distributed in phase-wrapped orbits throughout the large galaxy, and the resulting shells appear as density maxima at the instantaneous turnaround points of the star orbits.  Consequently, shells are often observed to be radially interleaved.  For example, the elliptical galaxy NGC 5128 contains a complex shell system thought to have originated from a merging disk galaxy $\sim$$10^9$ yr ago \citep{malin83b}.  CDM simulations predict that infalling material is often aligned on one or two preferred axes, often forming filamentary structures (for example, \citealt{navarro04} and references therein).  These results support the formation of shells through the slow collision between a large galaxy and a smaller companion.  Smaller systems are generally bluer than the large elliptical primaries, hence a characteristic of shells is the presence of a bluer population of stars compared to the host galaxy \citep{pence86,mcgaugh90}.  Further, the companion often contains a reservior of neutral gas which is shocked during the interaction, thereby producing a new population of stars.  This adds to the blue nature of the resulting shells, which may also retain complexes of H{\sc i} (for example, NGC 5128; \citealt{schim94}).  For the remainder of this section, we discuss alternative scenarios which may have resulted in the observed extra-tidal structure, and present the evidence supporting shell structure in Fornax.

\subsection{Alternative Scenarios}
If the overdense feature located $1.3^{\circ}$ to the NW of Fornax is a shell, then this would support our previous interpretation of the small structure located near the centre of the system \citep{coleman04a}.  However, it is worthwhile investigating other possible origins of the current feature.  An initially plausible explanation is that the feature represents a disrupted globular cluster.  Five globular clusters are known to be associated with Fornax \citep{hodge61a}, and they are approximately coeval with the formation epoch of the dSph \citep{buon98,buon99}.  The subtracted CMD we produced for the overdense region (Fig.\ \ref{cmdregsub}) does not preclude a star system of this age.  In addition, having subtracted the background stellar population, we measured the absolute integrated magnitude of the observed stars in this feature to be $M_V=-6.7 \pm 0.2$, which falls in the range defined by the other five Fornax clusters \citep{mackey03}.  It thus seems reasonable that this feature may be a sixth Fornax cluster, currently suffering a tidal pertubation due to the gravitational influence of its host galaxy.  However, the angular separation of this feature from the centre of Fornax is approximately $1.3^{\circ}$, while the globular clusters are all located within $0.67^{\circ}$.  \citet{rodgers94} found that the Fornax clusters $3-5$ show no evidence for tidal truncation, demonstrating the weak tidal field of Fornax.  Also, \citet{mackey03} did not find any sign of structures such as tidal tails in the five Fornax clusters.  It therefore seems unlikely that a distant cluster would display signs of extreme tidal distortion without a similar effect observed in the other clusters, unless the purported `sixth cluster' is contrived to be a very low density system on a highly elliptical orbit with a small pericentric passage.  Consequently, the likelihood of this feature being a disrupted globular cluster appears improbable.

A second argument against the disrupted globular cluster explanation is the presence of low surface brightness stellar structures discovered in the NW/SE extra-tidal regions of Fornax.  Both these features and the NW overdensity are abnormal in dSph galaxies, making it far simpler to assume they have a common basis.  A visual examination of the contour plot (Fig.\ \ref{fornaxcontour}) supports this view; these structures display the same orientation as the NW feature (along the minor axis) thus implying they stem from the same event.  Assuming they have a stellar population similar to the Fornax system, we estimate the lobed structures have a mean surface brightness (after background subtraction) of $\sim$$31$ mag/arcsec${}^2$, with an absolute integrated magnitude (to the survey limit) of $M_V = -8.8 \pm 0.4$.  We note that the structures may continue beyond the current survey limit, thus they may possess a higher luminosity than has been detected in the current survey.  However, this value is similar to the integrated magnitudes of the faintest Galactic dSphs, such as Ursa Minor and Draco.  

Another possibility is that these structures may be tidal tails, made up of stars stripped from Fornax by the gravitational field of the Galaxy.  Stripped stars follow the orbit of the satellite, and can fall back into this system at later times when the gravitational influence of the large host galaxy is reduced.  The Fornax structures appear similar to the tails formed in numerical models of a disrupting satellite orbiting a large host galaxy \citep{helmi01,mayer01,johnston02}.  These simulations also predict that the tidal tails may be clumpy, as matter is preferentially stripped from the satellite at perigalacticon.  Observations of disrupting satellites, such as Palomar 5 \citep{odenkirchen01} and the Sagittarius dwarf galaxy (\citealt{ibata94}; \citealt{majewski03} and references therein) support these predictions.  Hence, the NW feature may represent a clump in the tidal tail of Fornax.

However, clear tidal tails have only been observed in a single dSph satellite of the Galaxy; the Sagittarius dSph.  The Sagittarius orbit brings it within approximately 15 kpc from the Galactic centre with an orbital eccentricity of $e \approx 0.75$ \citep{law04}, thus experiencing substantial tidal forces.  In contrast, Fornax is located at a distance of 140 kpc \citep{m98}, and proper motion measurements indicate it is on a polar orbit with a small eccentricity ($e=0.27$), and is currently at perigalacticon \citep{dinescu04}.  Also, \citet{dinescu04} measured Fornax to be moving towards the ENE direction, which does not support tidal extensions to the NW/SE quadrants.  Moreover, Fornax contains a mass larger than all the other Galactic dSphs combined (excluding Sagittarius), which further precludes the possibility of stars escaping from this relatively deep potential well.  The Ursa Minor dSph also displays signs of interaction with the Galaxy; its internal structure is clearly disturbed \citep{olsz85,ih95,demers95,kleyna98} and UMi has a significant population of stars beyond the tidal radius \citep{mart-del01,palma03}.  However, this object lies at a Galactocentric distance less than half that of Fornax, and its total mass is approximately one third that of Fornax.  That is, UMi experiences stronger tidal forces than Fornax, and is less able to resist the consequential mass loss.  Therefore, we do not expect to see a more prominent effect of Galaxy-satellite interaction in Fornax compared to UMi.  Thus, given the relatively large mass, distance and orbital parameters of Fornax, the likelihood of it forming extra-tidal structures due to the Galactic tidal field is low.

\subsection{Evidence of Shell Structure}
Based on data from the deep Fornax survey by Stetson et al.\ (1998), we presented an analysis of an overdensity of stars located approximately $17'$ from the centre of Fornax, or just outside the core radius \citep{coleman04a}.  The overdense region of stars was shell-like in appearance, and after performing a similar CMD subtraction technique as described in \S \ref{nwshell}, we found it to be dominated by stars with an age $\sim$2 Gyr.  Our inference was that this feature represented shell structure in Fornax, the remnant of a collision between the dSph and a gas-rich companion approximately 2 Gyr ago.  However, this conclusion was based on a single shell with an absolute integrated magnitude of $M_V \sim -4$.  Here we have shown the presence of another shell-like feature, $M_V \sim -7$, located on the opposite side of Fornax at an angular separation of $80'$ from the centre of the dSph.  The two shells are concentric; they are aligned with the major axis of Fornax and are situated on the minor axis.  From these results, and the arguments presented above, we propose that this NW feature is a second shell, supporting the hypothesis of shell structure in a dSph.  In this case, the two lobes of material in the NW/SE extra-tidal regions represent those stars which are not at the extremes of their radial orbits.  This effect is seen in large galaxies containing a shell system; the region inside the shells display a small increase in surface brightness compared to the background \citep{malin83,mcgaugh90}.  Although we do not have deep enough photometry here to accurately investigate the stellar populations of these features, a supporting result for shell structure would be a prominent young stellar population similar that seen in the inner Fornax shell \citep{coleman04a}.

If a gas-rich companion merged with Fornax, then this encounter should have affected the star formation and chemical enrichment histories.  Fornax is known to have experienced a complex star formation history.  The deep photometric studies by Stetson et al.\ (1998) and Saviane et al.\ (2000) revealed that this system contains a young main sequence incorporating stars possibly as young as $10^8$ yrs, which is unusual compared to the other dwarf spheroidals associated with the Galaxy.  Indeed, \citet{pont04} examined the chemical enrichment history of Fornax using infrared spectra of 117 stars located towards the centre, and they note that the ``metallicity distribution and AMR [age-metallicity relation] that we derive for Fornax make it a complex system, more reminiscent of the LMC or the Milky Way's disk than of the other dSphs.''  Perhaps the most pertinent point relating to our encounter hypothesis was found by \citet{pont04}, who were able to reproduce the qualitative appearance of the Fornax CMD using their AMR.  They noted that Fornax contains a tail of relatively metal rich red giants, with an approximate abundance range of $-0.7 \le$ [Fe/H] $\le -0.4$, and that these stars have an age of $\sim$2 Gyr.  Indeed, \citet{pont04} have shown that Fornax experienced a substantial metallicity increase in the last few gigayears, and that the star formation rate increased by a factor of $\sim$2 in the last $2-4$ Gyr.  Although this result contains uncertainties depending on stellar evolution models, it is highly relevant to our claim of a merger between Fornax and a gas-rich companion $\sim$2 Gyr ago.  The ensuing gas shocking mechanisms would have resulted in a significantly accelerated star formation.

Shells are formed during the slow collision between a galaxy and a small counterpart, and this process is dependent on having a relatively fixed potential and a dynamically cold intruder.  As such, we do not expect the overall structure of Fornax to have been adversely affected by this event, and this is supported by our analysis of the inner regions of Fornax; apart from a slight radial dependence of isophote centre, position angle and ellipticity, there are no signs of significant perturbation.  The radial profile we derived for Fornax indicated an excess of stars compared to the King model at large radii, which is often interpreted as representing extra-tidal structure.  It is worth noting that \citet{dekel03} performed simulations describing the build of dark halos through merging satellites.  They predict that an infalling satellite with a ``puffy'' structure may cause baryonic feedback effects in the central regions of the host system, causing in an expanding halo centre and initiating a burst of star formation.  The resulting radial profile resembled a power law distribution with a flat core, qualitively similar to the profile we derived for the Fornax RGB stars.

Stetson et al.\ (1998) note that the most recent burst of star formation appears to have been concentrated at the centre of Fornax and the young main sequence stars define a bar ($\sim$20 arcmin in length) possibly with two lobes at either end (see their Fig.\ 12).  This distribution bears a remarkable similarity to the `bow-tie' structures produced in the shell formation models of \citet{hernquist88,hernquist89}.  The concentric nature of the two Fornax shells, and their location on opposite sides of Fornax, support a low angular momentum (radial) encounter.  On the other hand, the young MS bar found by Stetson et al.\ (1998) is not aligned with these shells, which, on the basis of the Hernquist \& Quinn models, would suggest a high angular momentum (non-radial) encounter.  However, a high angular momentum encounter is often characterised by chaotic streams which we have not detected in Fornax, and thus this particular encounter may fall somewhere between the two scenarios.  Modelling of the interaction is required to shed further light on the merger process.

Large galaxies have been observed to contain up to 20 shells \citep{malin83} which are interleaved with radius.  Consequently, Fornax may contain a larger number of shells than has been observed here, and these may also be radially interleaved.  Shells occur at the turnaround points of the stellar radial orbits, and hence those at larger radii contain stars with longer orbital periods, thus implying a longer formation time than those at small radii.  That is, the formation of shells propagates outwards with time.  As such, a deeper wide-field study of Fornax may reveal more shells closer to and further from the galaxy centre than the NW feature.  Also, stars occupying the shell are turning around in their orbits, and hence the motion of these stars (relative to the host galaxy) will be close to zero \citep{merrifield98}.  Consequently, shells are expected to be kinematically cold, and the radial velocity dispersion of the Fornax shells may further reveal the nature of this interaction.  Dwarf galaxies are dominated by dark matter, hence numerical simulations (with a complete model of baryonic and dark matter) are required to confirm if the slow collision of two dwarf galaxies can produce the structure observed in Fornax.  We note though that the extra-tidal structures appear remarkably similar to the models of interacting dark halos presented by \citet{knebe04}.

\section{Conclusion}
Accurate photometry in two colours ($V$ and $I$) was obtained for a 10 deg${}^2$ region centred on the Fornax dSph galaxy with the aim of searching for extra-tidal structure.  We have complete photometry down to a magnitude of $V=20$, which corresponds to stars approximately 1 mag brighter than the Fornax red clump.  Stars were selected using the CMD of Fornax as a mask, thereby increasing the contrast of Fornax red giants compared to the field star population.  The radial profile revealed a significant excess of stars compared to the best-fitting King profile beyond a major axis radius of $\sim$$50'$, a strong indication of extra-tidal structure.  The distribution of candidate RGB stars revealed a high-density feature located $1.3$ deg NW from the centre of Fornax, or approximately 30 arcmin beyond the nominal tidal radius.  The feature has a similar appearance to the shells discovered in E galaxies \citep{malin80,malin83,schweizer80}, and is located on the opposite side of Fornax to the original shell postulated by \citet{coleman04a}.  We removed the background stellar population, and calculated the absolute integrated magnitude from all observed stars in the NW feature to be $M_V = -6.7 \pm 0.2$.  After including the unseen light from stars below the survey limit ($\sim$$30\%$ of the total, depending on the IMF and age of the stellar population) then we estimate the total light from the NW feature to be $M_V \sim -7$.

A full statistical analysis of the density of candidate red giants in the extra-tidal region revealed two `lobes' to either side of Fornax, located in the NW and SE quadrants.  These lobes are situated along the minor axis, symmetric about the Fornax major axis, and share the alignment of the two shells mentioned above.  After removing the background population (and assuming these features have approximately the same luminosity function as Fornax) we measured their mean visual surface brightness to be $\mu_V \sim 31$ mag/arcsec${}^2$.  Thus, to the spatial limit of our survey, they display an integrated absolute magnitude of $M_V = -8.8 \pm 0.4$, approximately equal to that of the faintest Galactic dSph systems in Ursa Minor and Draco \citep{m98}.  We propose these structures and the new NW shell are the remnant of an merged companion galaxy, the same event we claim to have caused the inner shell \citep{coleman04a}.

In addition to the original Fornax shell, these structures present strong evidence for an interaction with a smaller companion in the relatively recent past.  CDM simulations predict that the accretion process is scale-free, hence dwarf galaxies are predicted to have formed through the merger of several smaller systems.  Although recent infall is not expected to have occurred in dwarf galaxies, our observations provide evidence for a recent accretion into the Fornax system.  Albeit at a smaller scale, this scenario is analogous to the merger event occurring between the Sagittarius dwarf galaxy and the Milky Way.  Deep photometric (and spectroscopic) data are required to explore the nature of these Fornax structures.  If the stellar population is similar to the Fornax system, this would indicate the substructures are tidal tails.  However, if the stars in the shell display a marked difference from those in Fornax (such as a blue, kinematically cold population), this would support the possibility of shell structure in a dSph.

\acknowledgments
The authors acknowledge Brian Schmidt for providing the astrometric calibration program and Agris Kalnajs for his advice on galaxy interactions.  MC thanks Paul Allen for his helpful discussions on the angular correlation function.  MC acknowledges the financial support provided by an Australian Postgraduate Award.  This research has been supported in part by funding from the Australian Research Council under Discovery Project Grant DP034350.


\cleardoublepage

\plotone{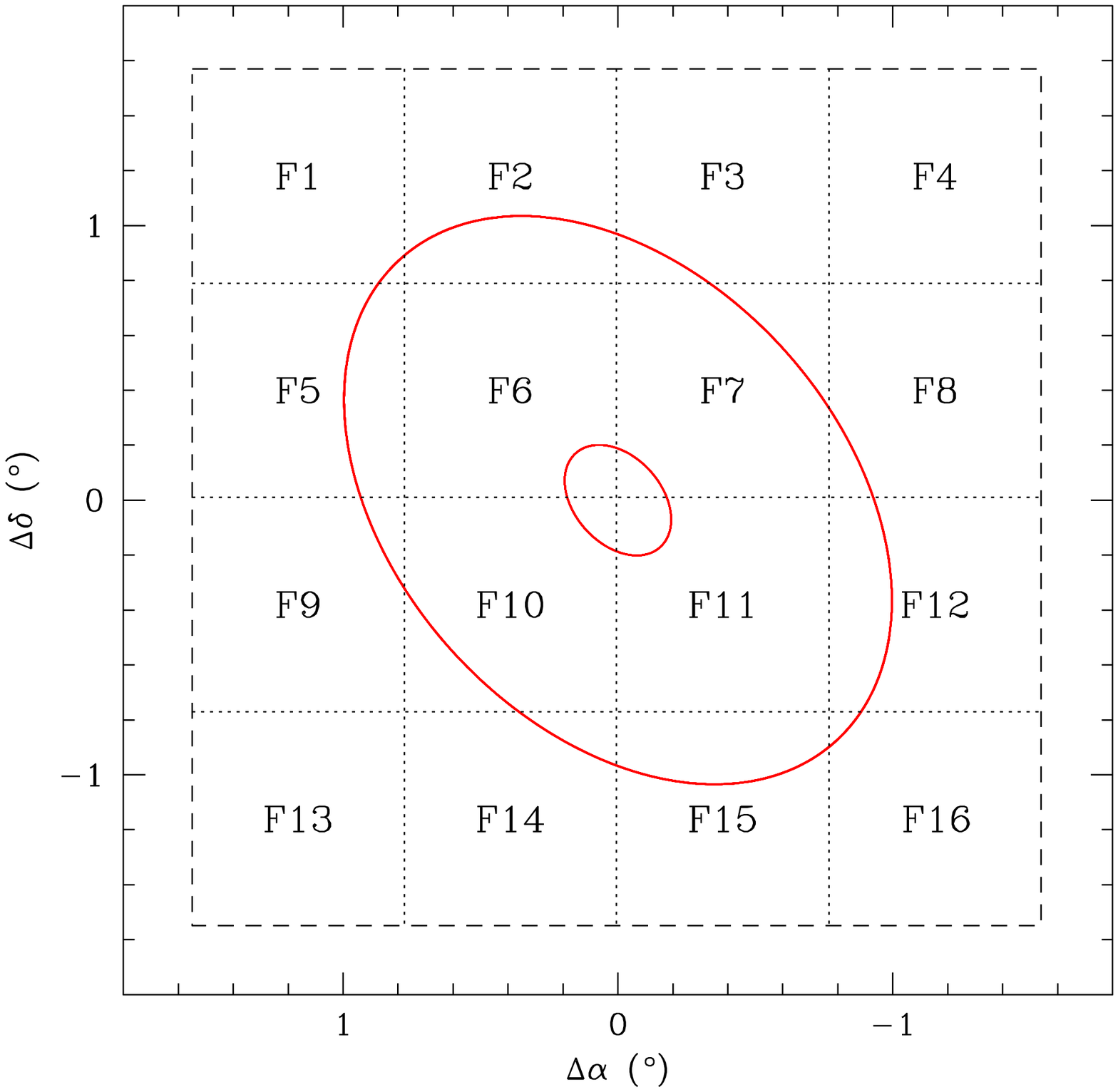}
\figcaption[Schematic map of observations]{Schematic diagram illustrating the region covered by this survey.  The red lines trace the core and tidal radii of Fornax \citep{m98} and outer dashed line represents the survey limit.  The `boundaries' between fields are represented by dotted lines, although each field contains an overlap region (width $\sim$$5'$) with all adjacent fields.  The fields are labelled $\mbox{F1}, \dots, \mbox{F16}$. \label{obsmap}}

\plotone{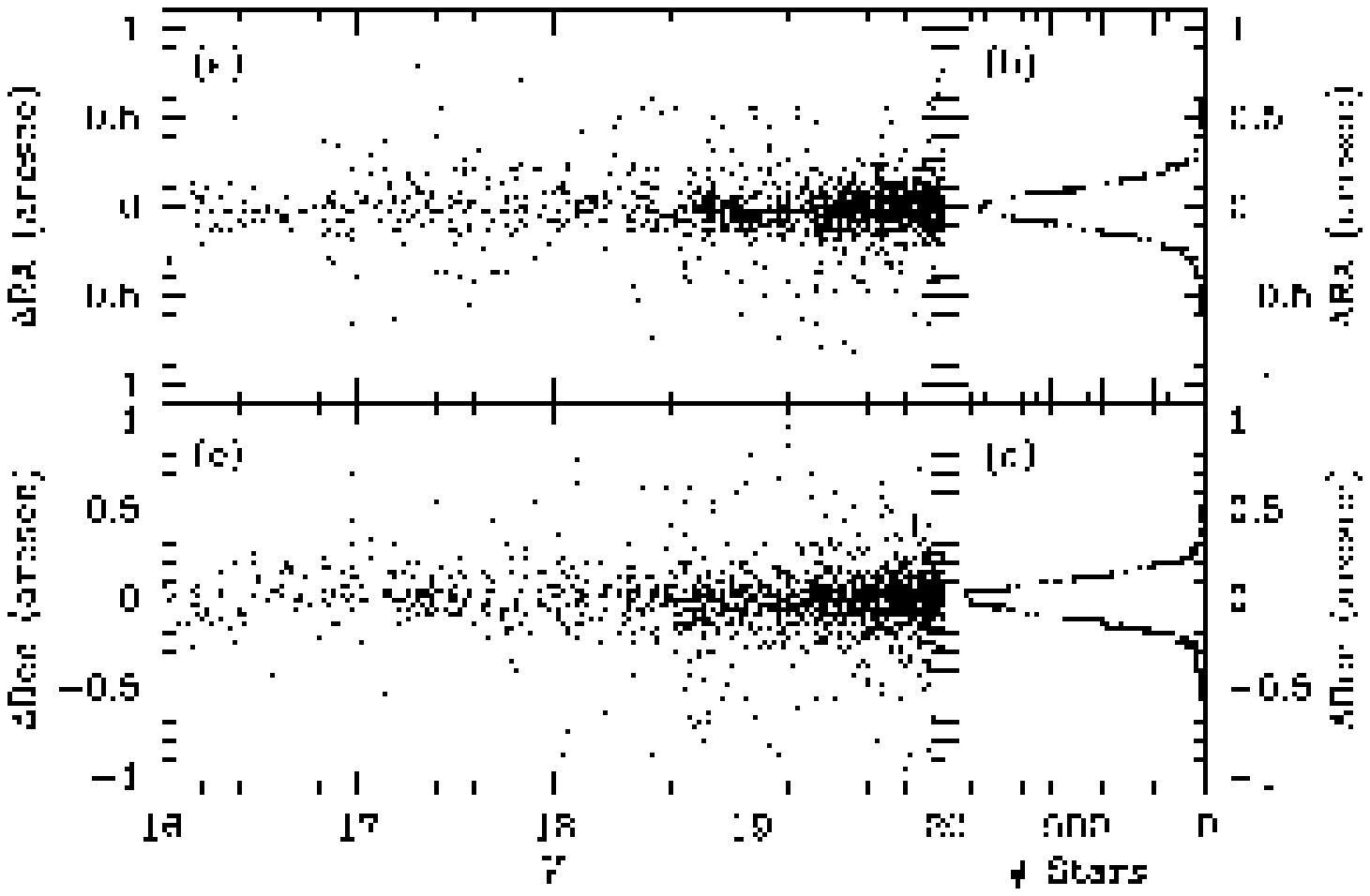}
\figcaption[Astrometry in $V$]{(a) RA differences of multiply detected objects in the range $16 \le V \le 20$.  (b) Frequency distribution of $\Delta$RA values.  The dotted line shows the best Gaussian fit, centred at $-0.0069''$ with a standard deviation of $0.1096''$.  (c) Same as (a) for $\Delta$Dec.  (d) Same as (b) for $\Delta$Dec.  This distribution is centred at $-0.0035''$ with a standard deviation of $0.1027''$.  \label{coordsV}}

\plotone{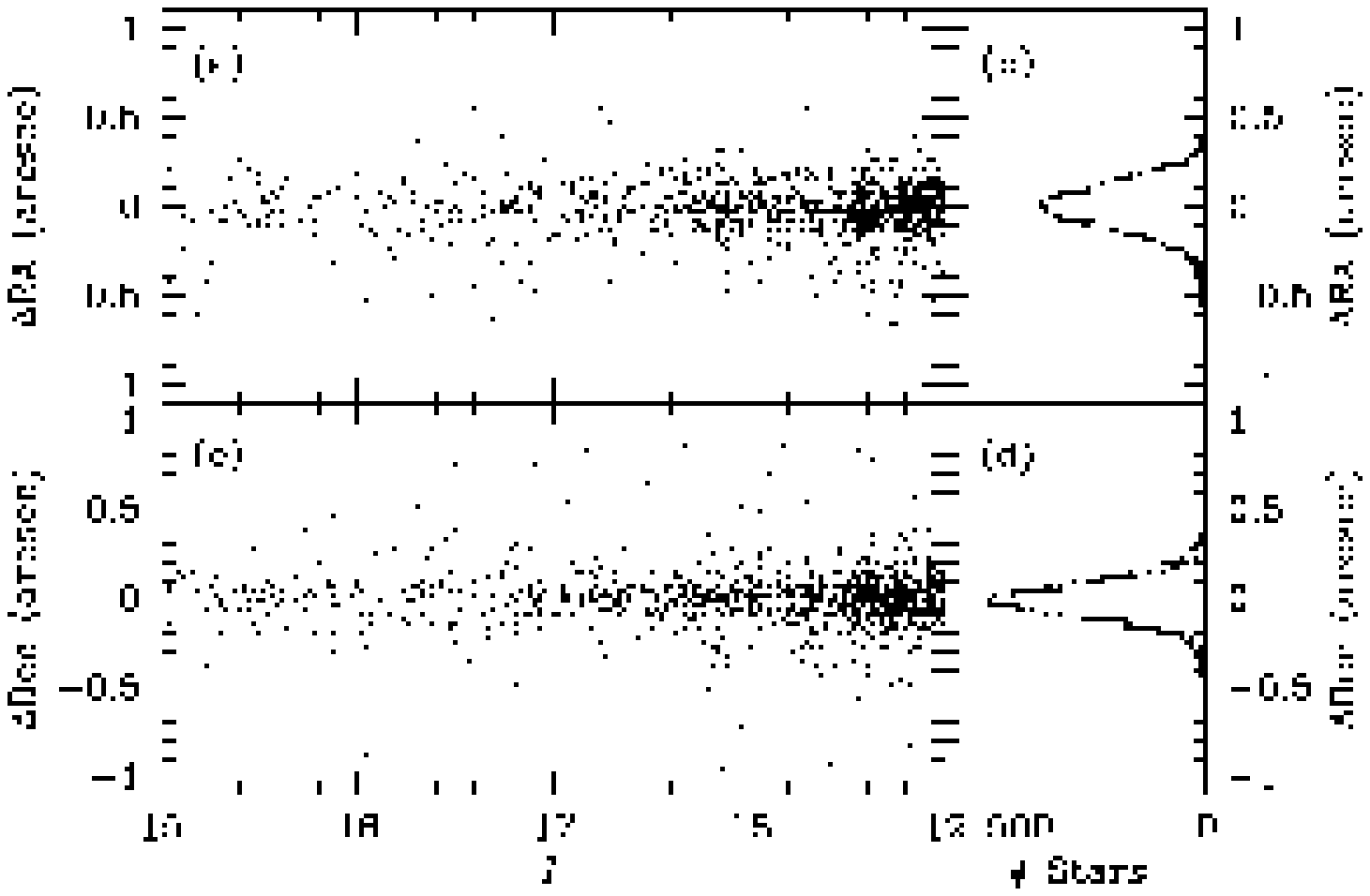}
\figcaption[Astrometry in $I$]{(a) RA differences of multiply detected objects in the range $15 \le I \le 19$.  (b) Frequency distribution of $\Delta$RA values.  The dotted line shows the best Gaussian fit, centred at $0.0036''$ with a standard deviation of $0.1294''$.  (c) Same as (a) for $\Delta$Dec.  (d) Same as (b) for $\Delta$Dec.  This distribution is centred at $-0.0107''$ with a standard deviation of $0.0933''$.  \label{coordsI}}

\plotone{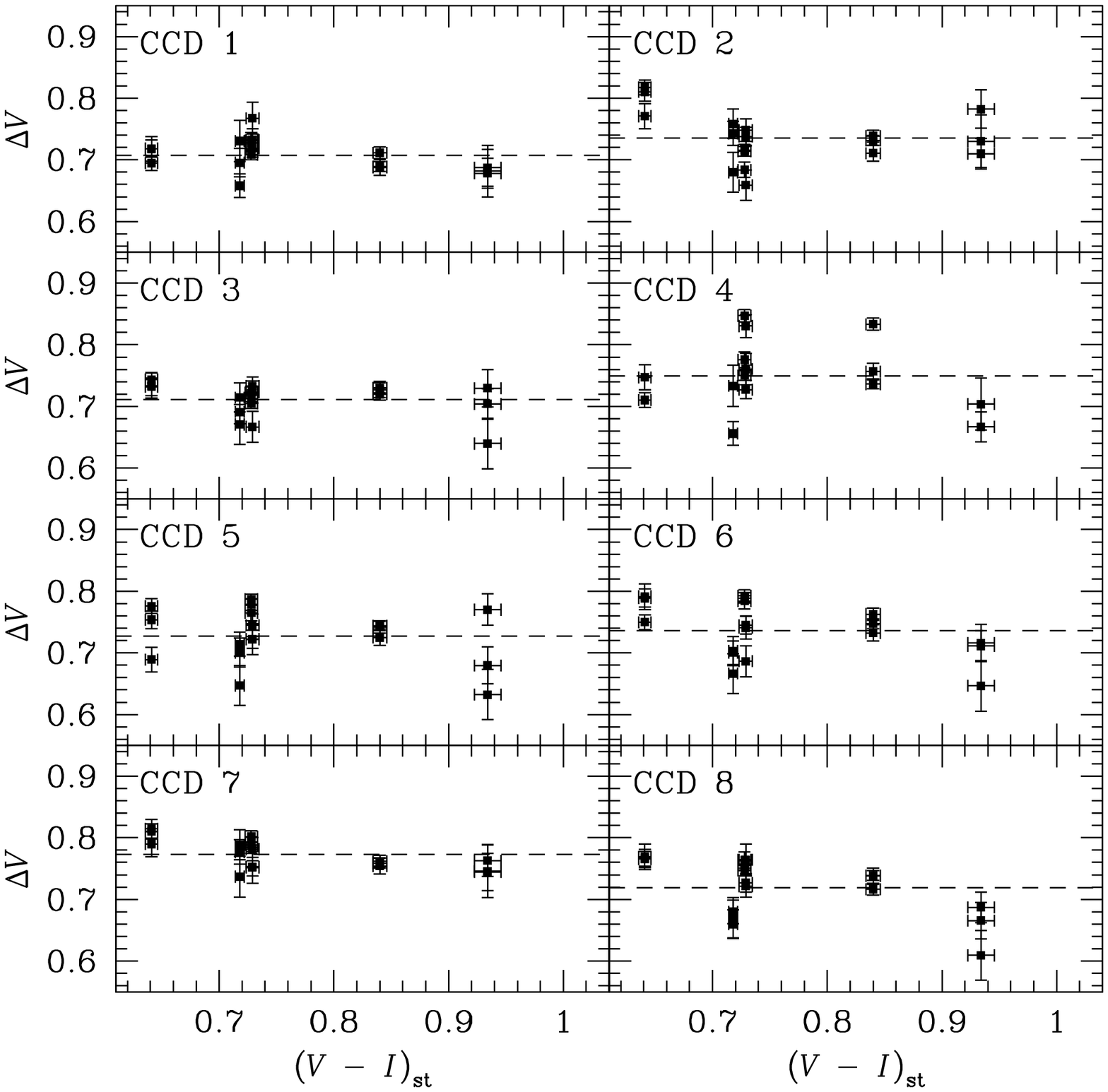}
\figcaption[Instrumental magnitudes over WFI CCDs - $V$]{Instrumental magnitude comparison between all WFI CCDs in $V$.  The dashed line represents the average of these data points. \label{Vstandards_ccd}}

\plotone{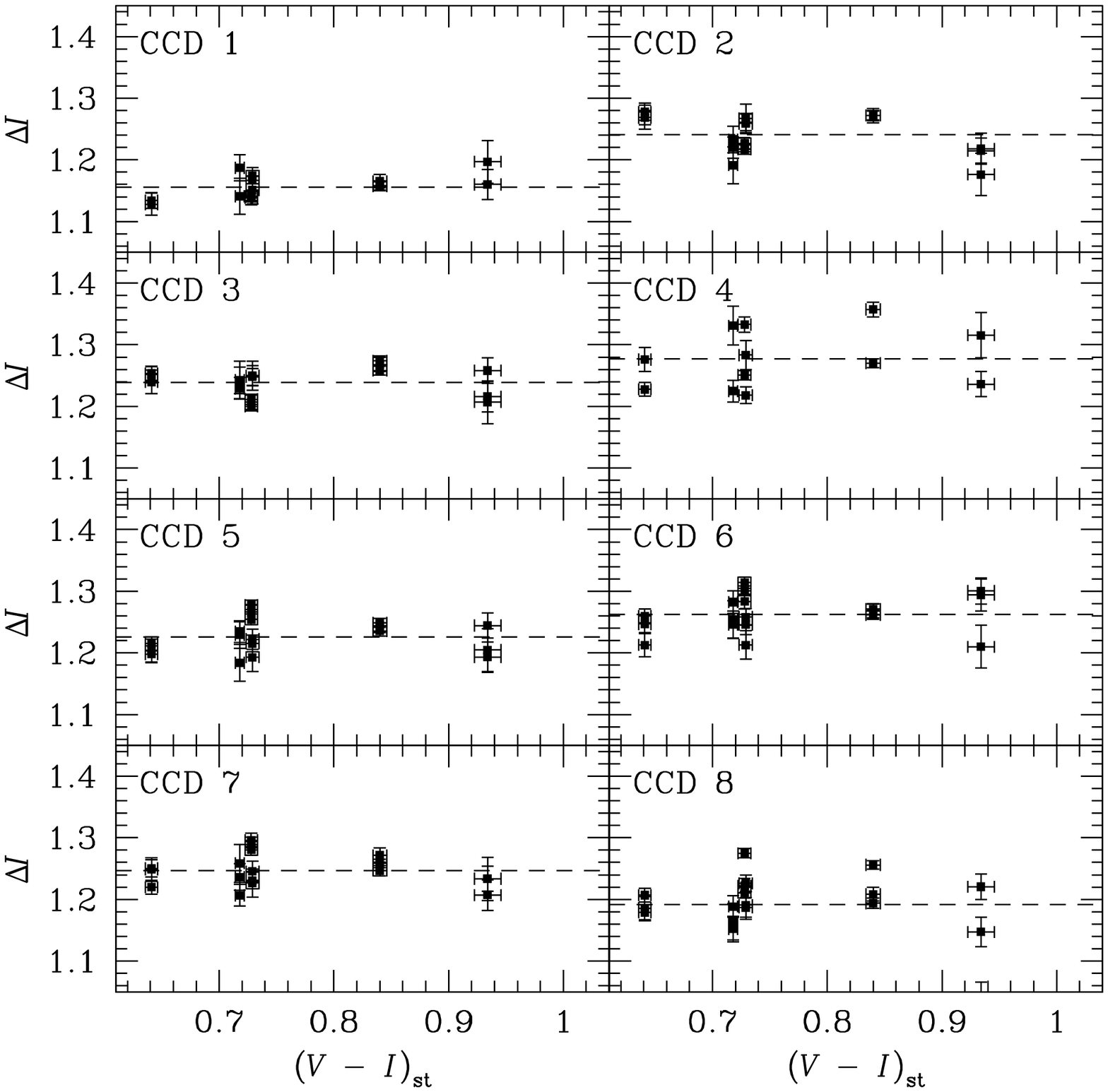}
\figcaption[Instrumental magnitudes over WFI CCDs - $I$]{Instrumental magnitude comparison between all WFI CCDs in $I$.  The dashed line represents the average of these data points. \label{Istandards_ccd}}

\plotone{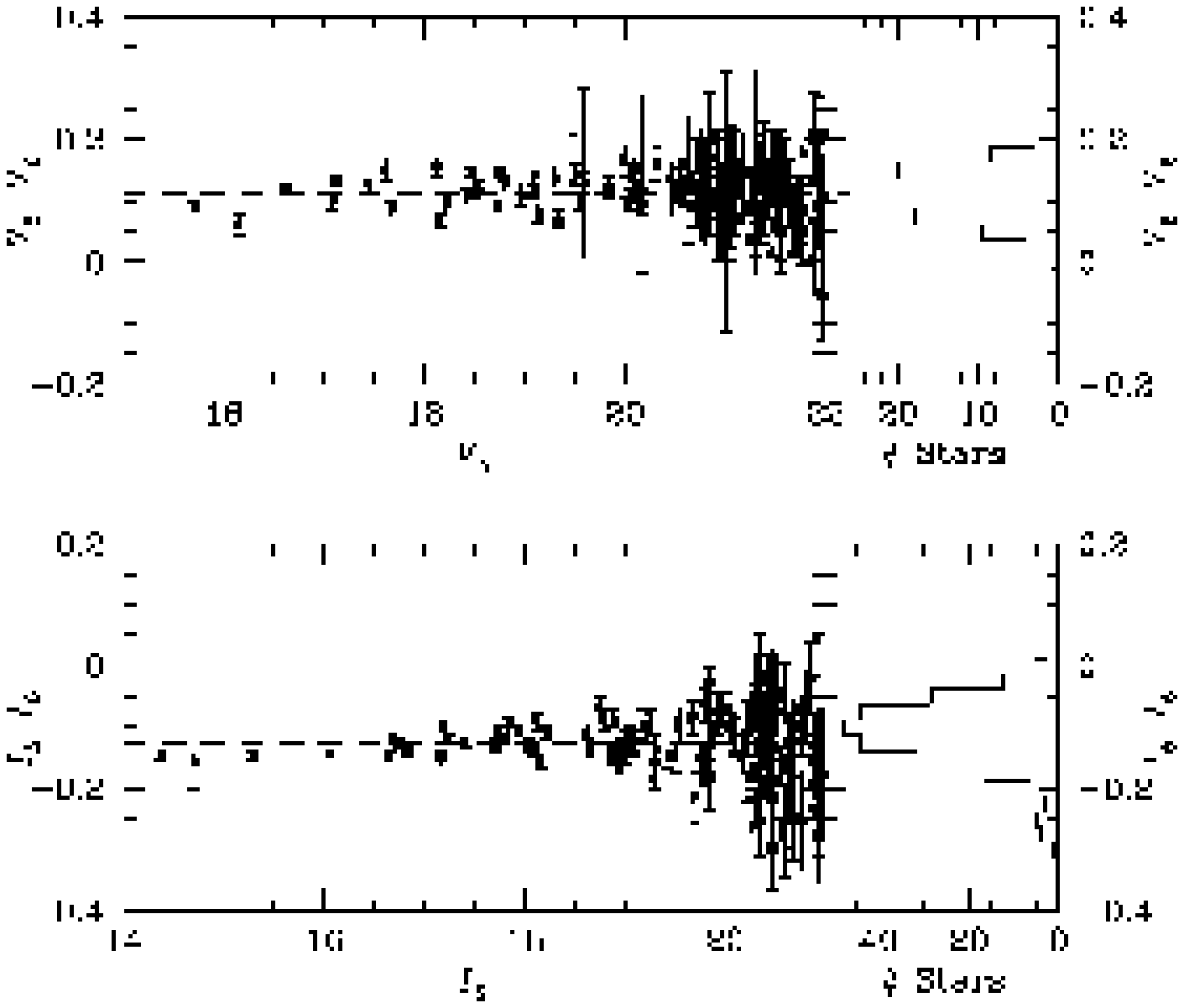}
\figcaption[Fields 5/6 calibration]{Difference in magnitude between matched stellar pairs in the overlap region of fields five and six.  The dashed line is the least-squares fit to these points.  To the right of each figure is the distributon of these magnitude differences, the standard deviation of which provided the differential magnitude error. \label{field5calib}}

\plotone{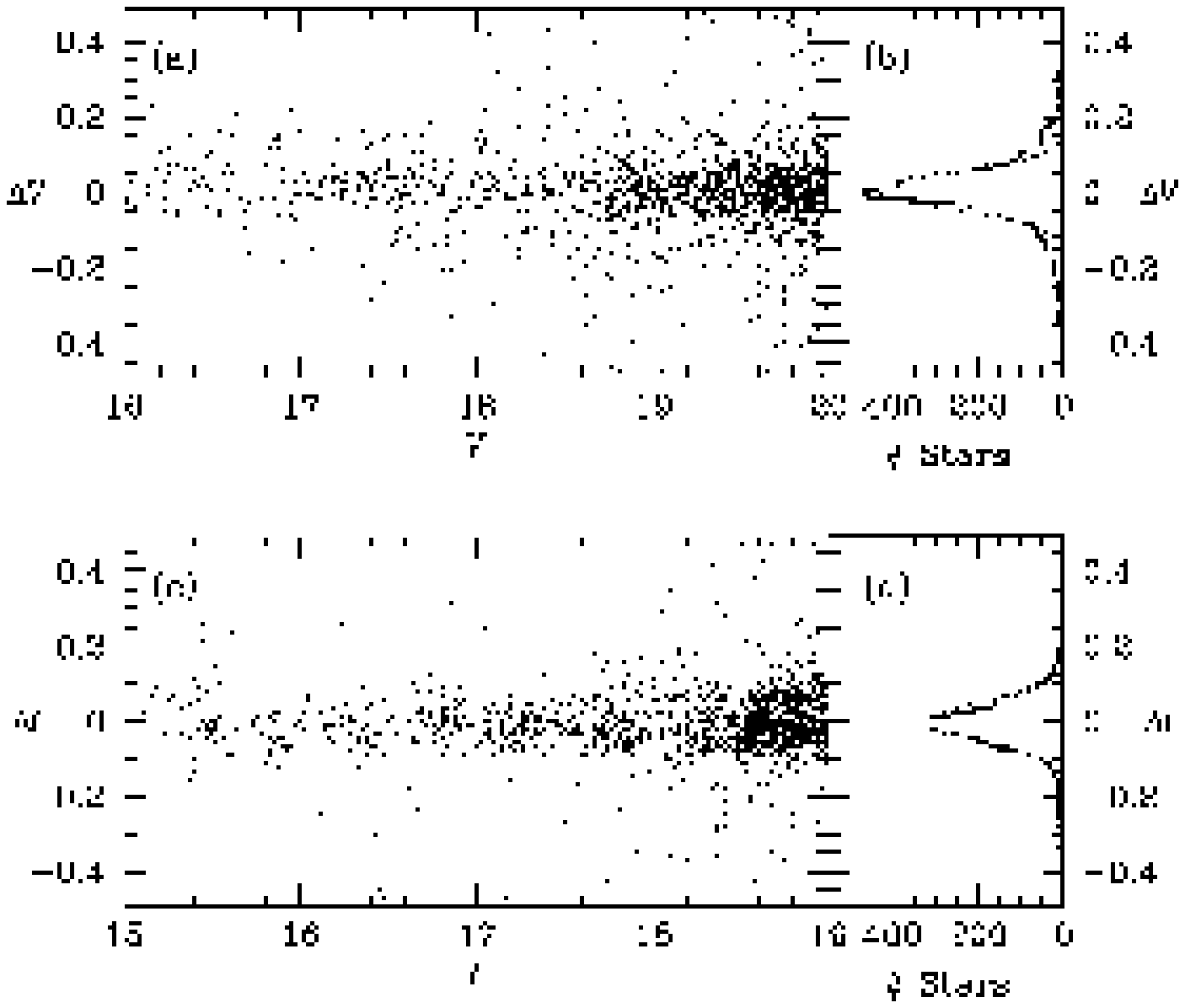}
\figcaption[Internal accuracy of the Fornax photometry]{(a) $V$ magnitude differences of all fields for multiply detected objects in the range $16 \le V \le 20$.  (b) Frequency distribution of $\Delta V$ values.  The dotted line shows the best Gaussian fit, centred at $\Delta V=0.005$ with a standard deviation of $0.048$ mag.  (c) Same as (a) for $\Delta I$ in the range $15 \le I \le 19$.  (d) Same as (b) for $\Delta I$.  This distribution is centred at $\Delta I = -0.001$ with a standard deviation of $0.051$ mag.  \label{phothist}}

\plotone{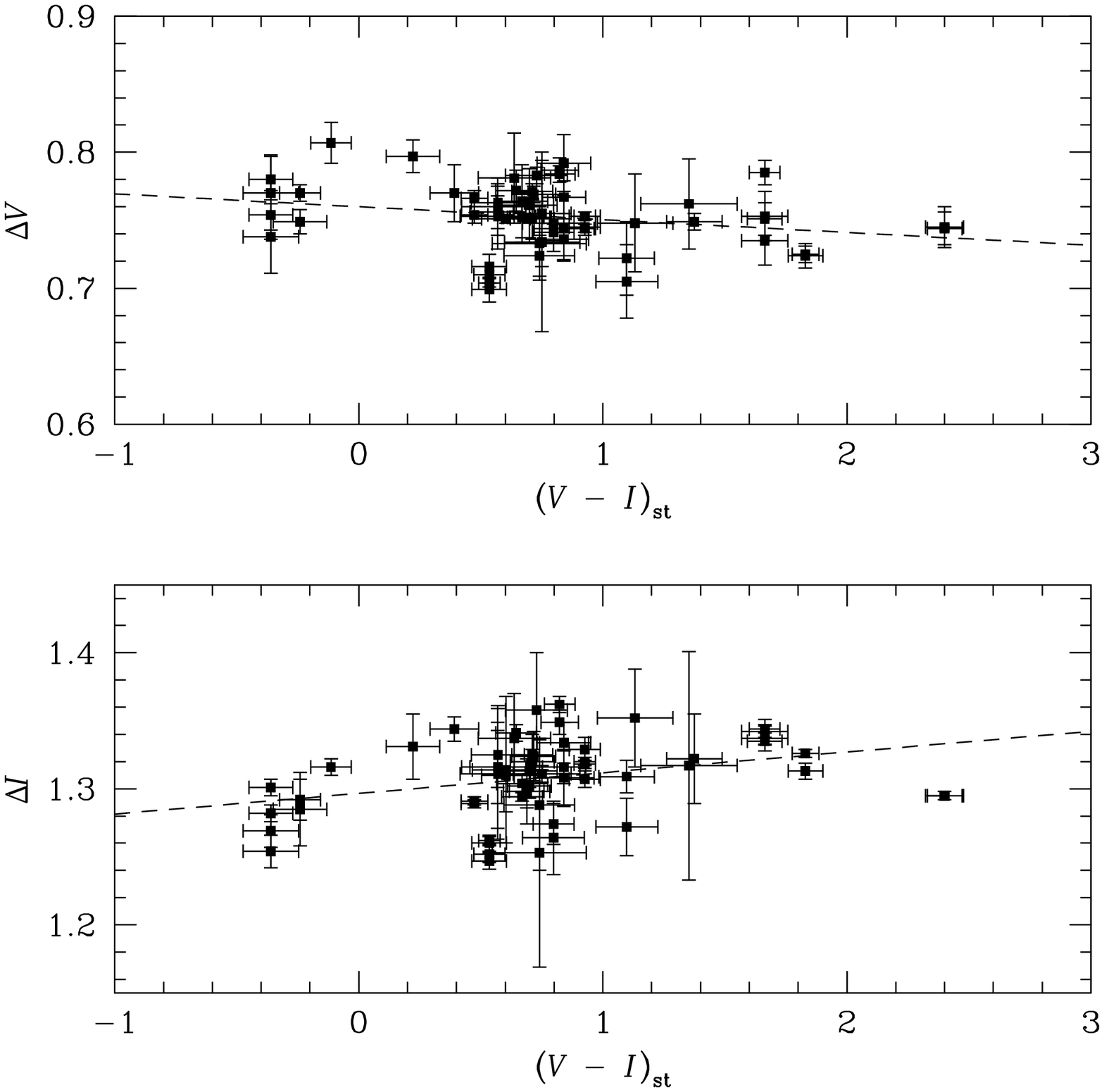}
\figcaption[Transformation equations solution -- 20th Oct 2001]{The difference between instrumental and standard magnitudes against colour for the standard field observations of 20th October 2001. \label{20octstds}}

\plotone{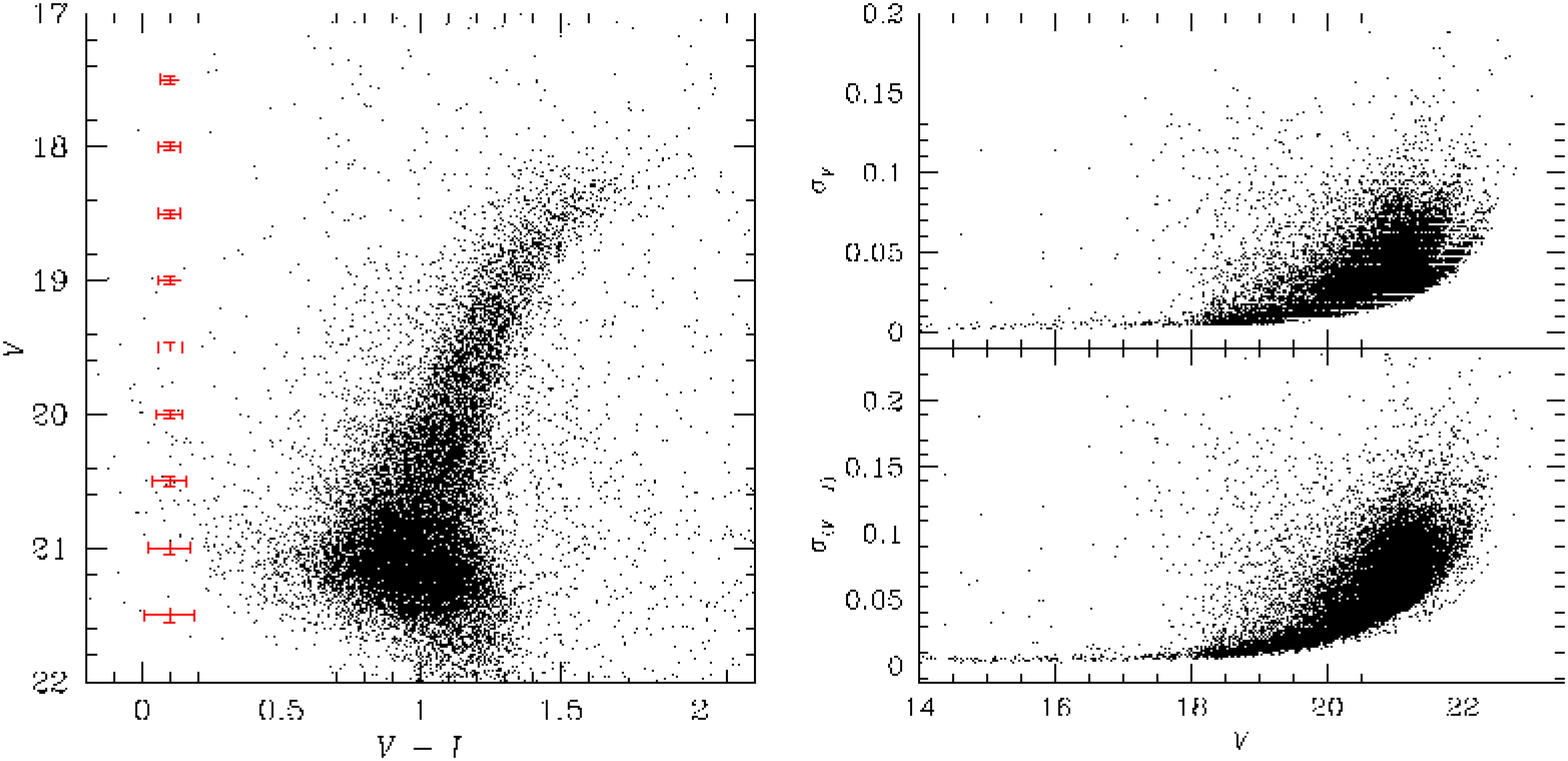}
\figcaption[Fornax CMD]{{\em Left panel:} Colour magnitude diagram from the inner $30'$ of Fornax.  The error bars to the left are the {\em rms}  of the DAOPHOT photometric errors ({\em right}).  {\em Right panel:} Corresponding DAOPHOT photometric errors, $\sigma_V$ and $\sigma_{V-I}$ as a function of $V$. \label{fornaxcmd}}

\plotone{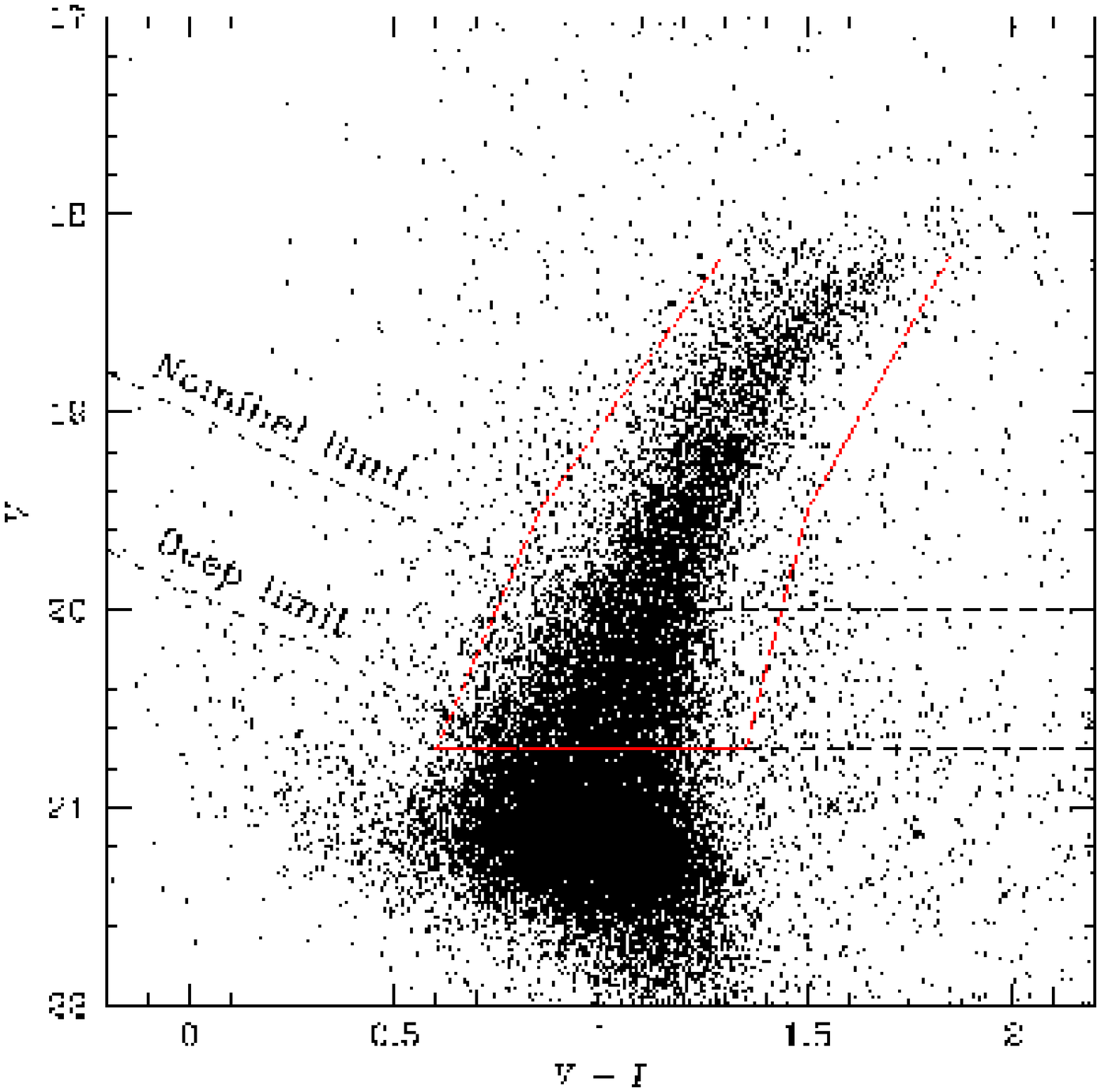}
\figcaption[Fornax CMD]{Colour-magnitude diagram for the inner $30'$ of Fornax.  The upper dotted line represents the photometric limit of the dataset, and the lower dotted line is the photometric limit imposed on the `deep' fields.  The solid line outlines the region to select candidate Fornax red giant stars. \label{fornaxcmdlines}} 

\plotone{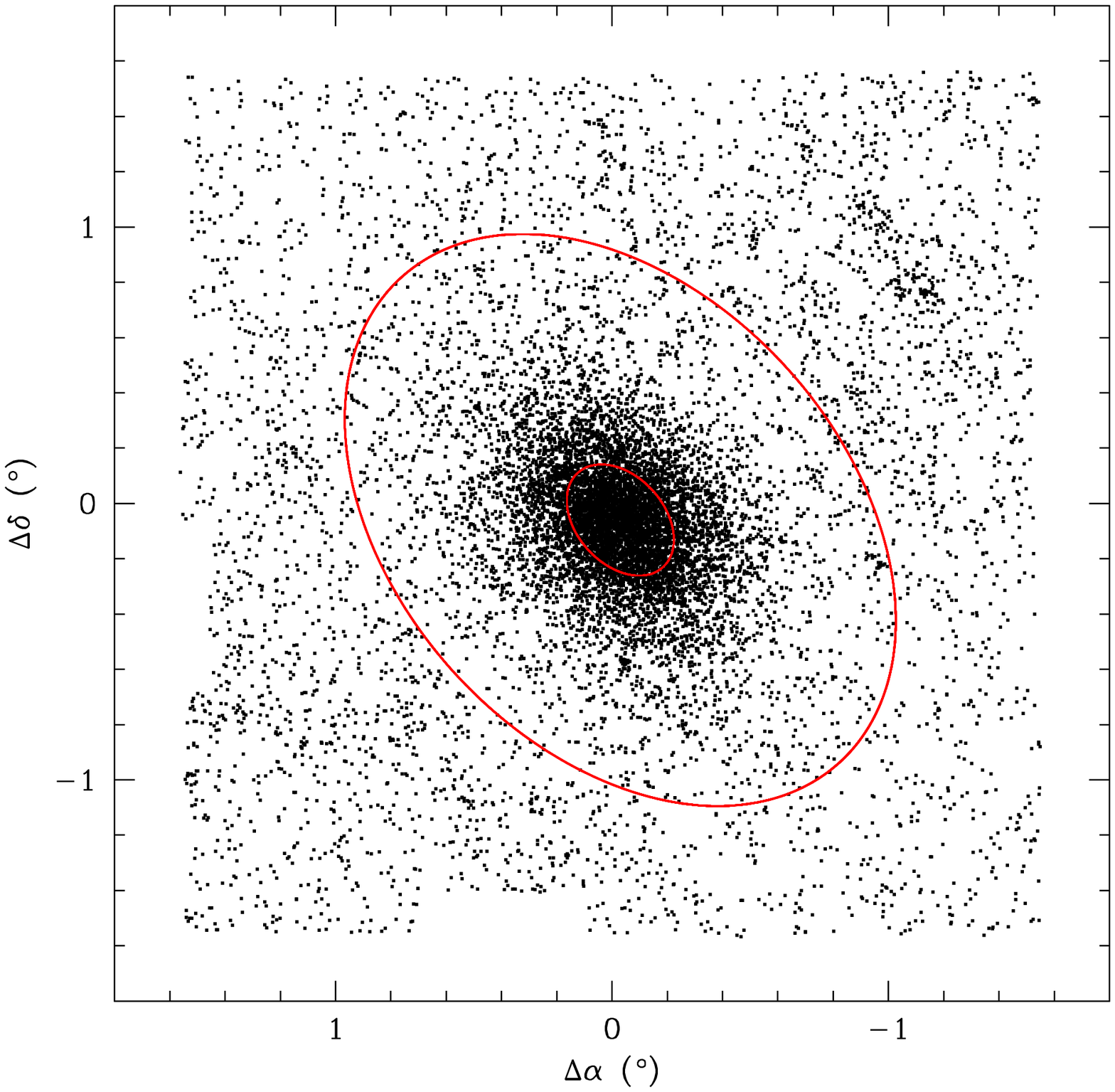}
\figcaption[Spatial distribution of Fornax RGB stars]{Distribution of RGB-selected stars on the sky.  The inner and outer ellipses are the core and tidal radii respectively \citep{m98}.  The original shell described in \citet{coleman04a} is located approximately $17'$ SE of the Fornax centre, slightly beyond the core radius. \label{fornaxrgbxy}}

\plotone{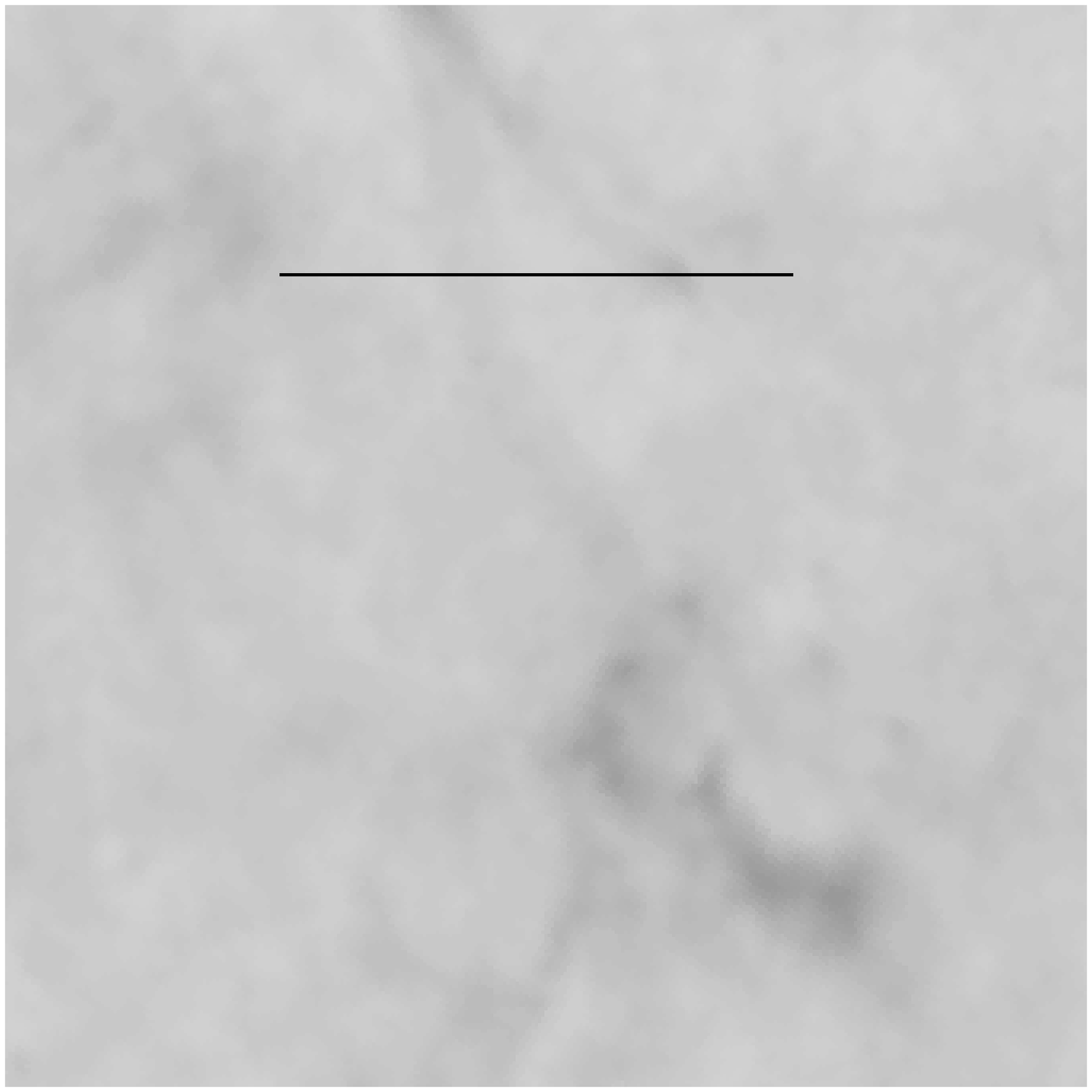}
\figcaption[Dust map in the Fornax region]{IRAS $100 \mu$m map in the Fornax region \citep{schlegel98}, where the darker shading reflects a higher dust density.  The black square outlines the $3.1 \times 3.1$ deg${}^2$ survey area.  North is up and East is to the left.  The majority of the Fornax region has a of reddening $E(B-V) \sim 0.02$, while the darker region towards the South-West of the survey region corresponds to a reddening of $\sim 0.07$. \label{fornax_dust}}

\plotone{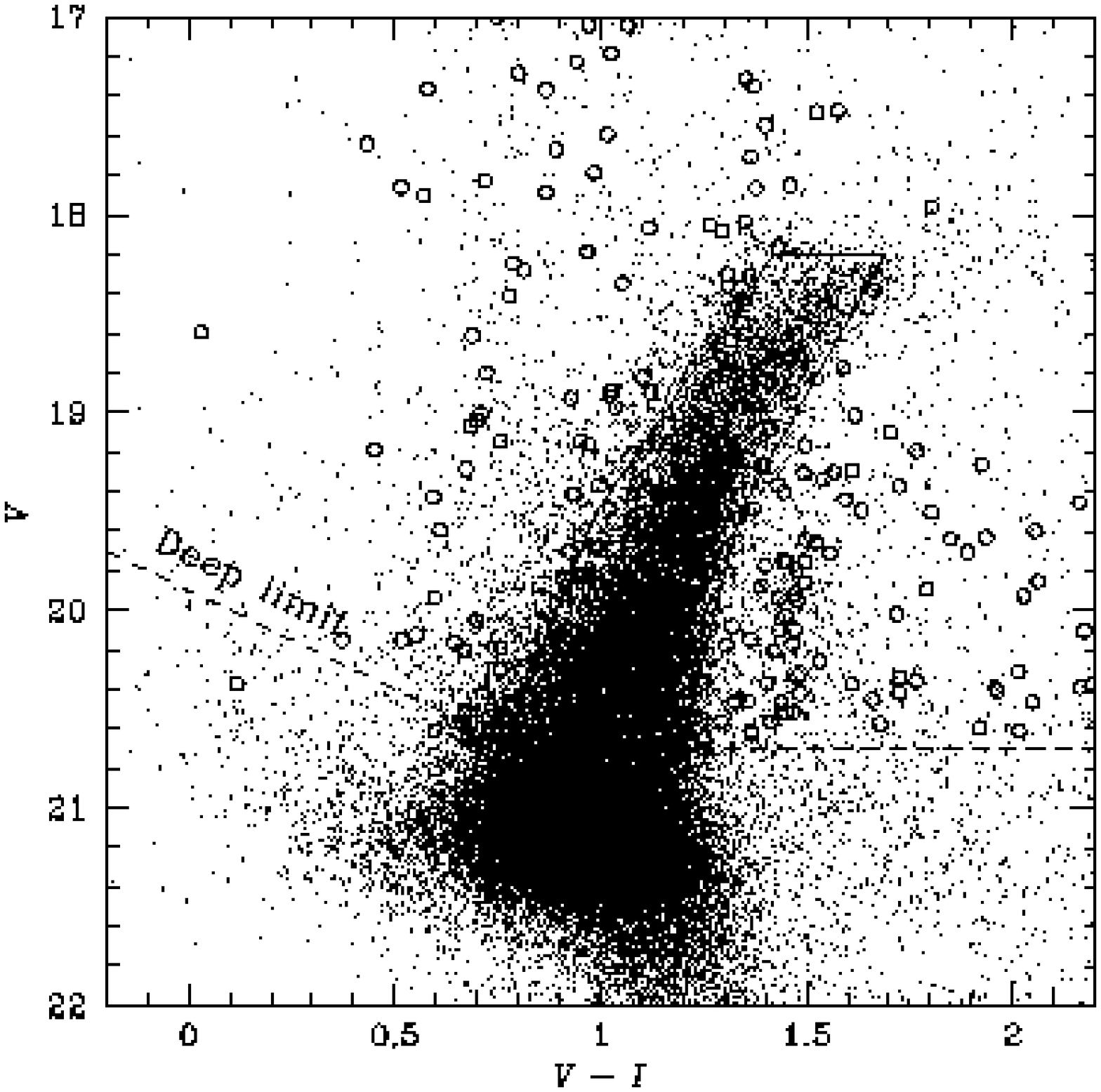}
\figcaption[Overdensity CMD]{The open circles represent stars situated in the overdense region, while the black points are the CMD data displayed in Fig.\ \ref{fornaxcmd} (with a smaller point size).  The red line outlines the optimised selection region for the stars displayed in Fig.\ \ref{fornaxrgbdeepxy}. \label{clumpcmd}}

\plotone{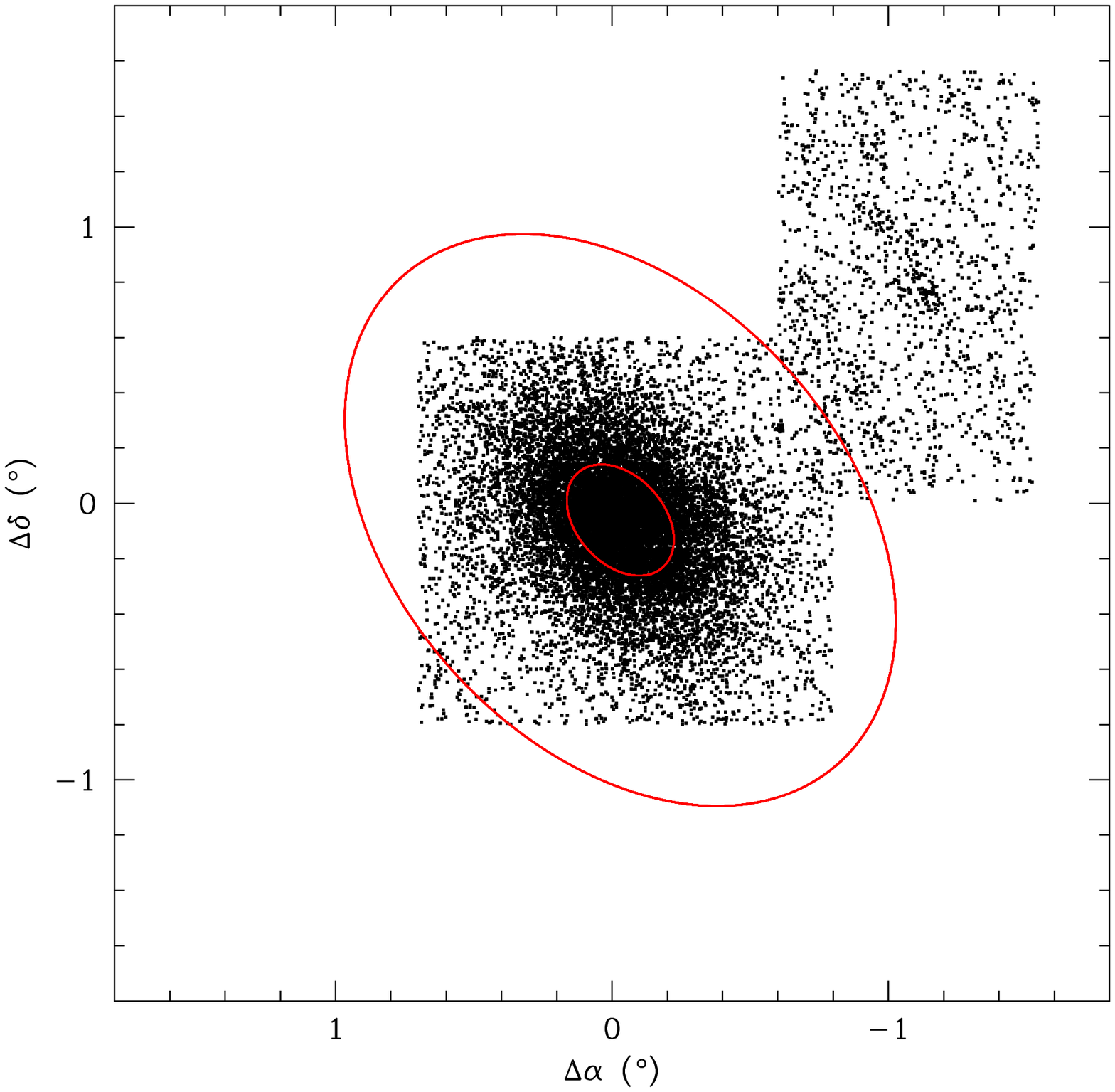}
\figcaption[Fornax RGB stars down to $V=20.7$]{Extending the RGB-selection down to $V=20.7$.  The selection region is shown in Fig.\ \ref{clumpcmd}, and has been optimised to increase the contrast of the overdense feature to the NW of Fornax.  Fields four, eight and the four inner fields are shown. \label{fornaxrgbdeepxy}}

\plotone{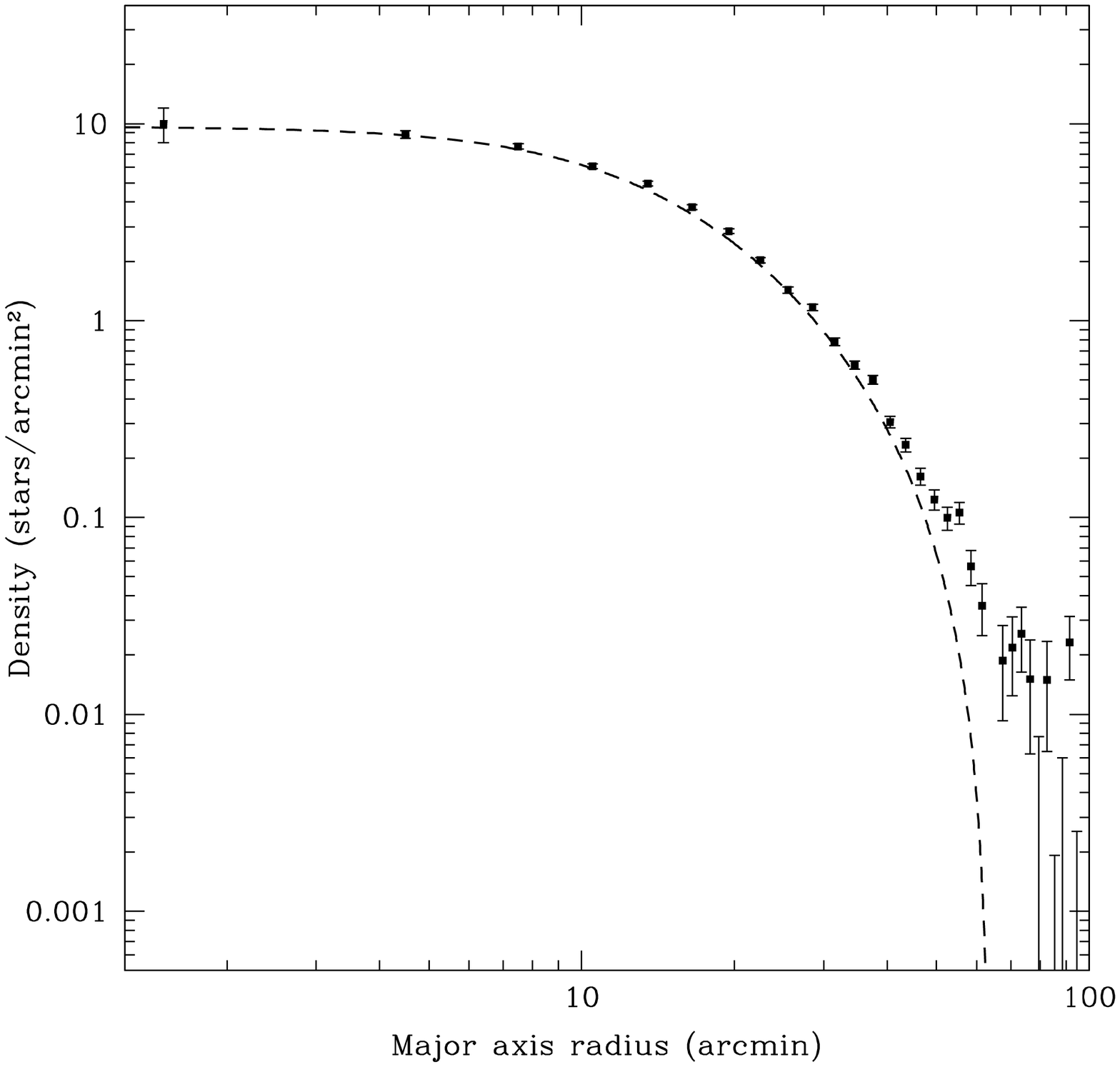}
\figcaption[Radial profile of Fornax RGB stars]{Radial profile of the Fornax RGB-selected stars, where the error bars are determined from Poisson noise.  A background level has been subtracted from all data points, calculated to be $0.141 \pm 0.013$ stars/arcmin${}^2$ from all density values beyond a major axis radius of $80'$.  The dashed line represents the best-fitting King model to all points, with a core and tidal radius of $r_c = 16.0'$ and $r_t = 63.9'$ respectively. \label{radial}}

\plotone{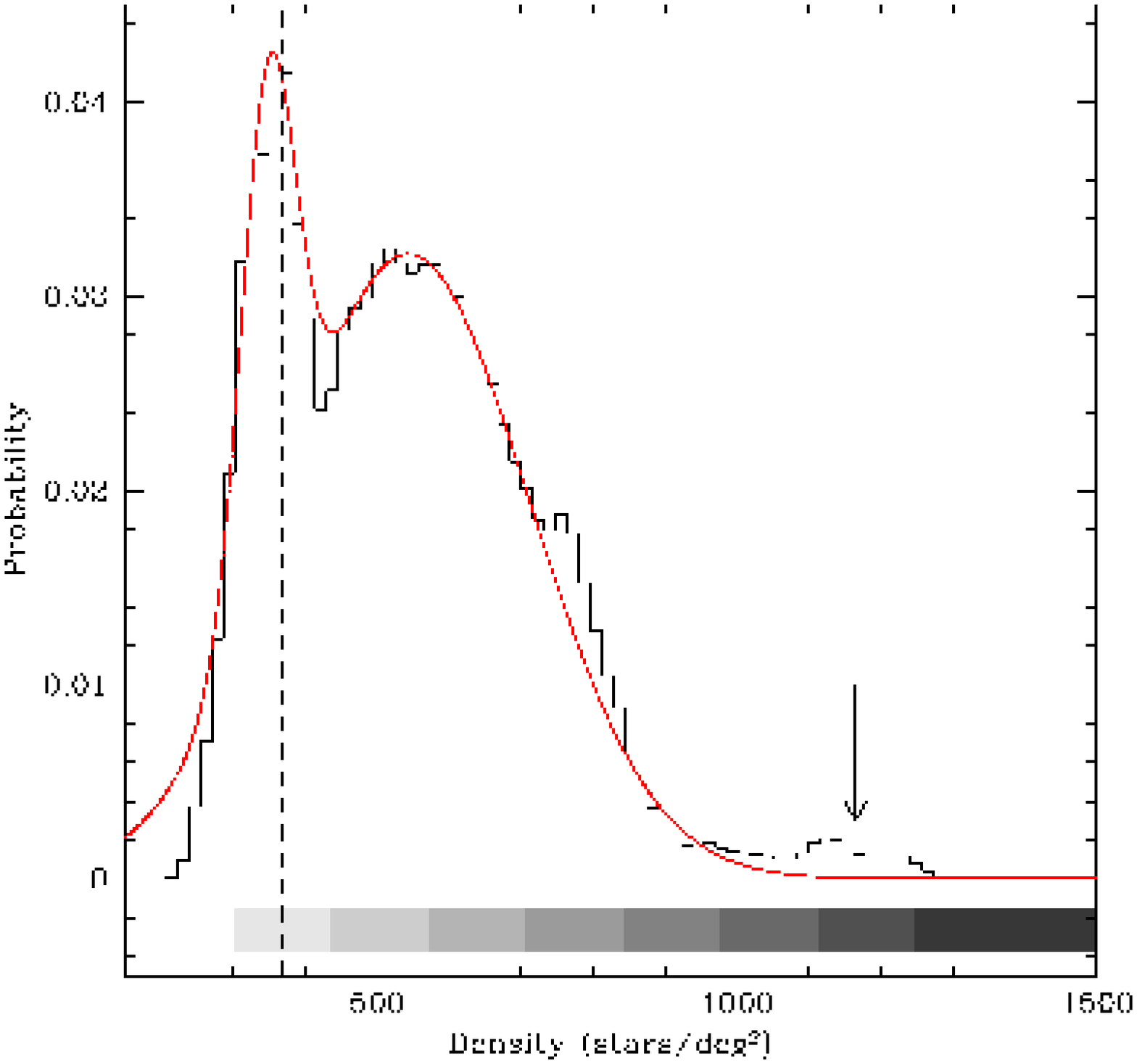}
\figcaption[Probability density function]{Monte Carlo evaluation of the density distribution of Fornax RGB-selected stars, showing the probability of measuring a given density (per circle of radius $12'$) beyond the nominal tidal radius.  The dashed line corresponds to the (normalised) \citet{rat85} star counts.  The red line indicates the sum of two best-fit Gaussians to the low and medium density background populations, and the arrow indicates the high density peak which corresponds to the NW overdensity.  The greyscale bar represents the density scale used for the contour plot in Fig.\ \ref{fornaxcontour_outer}. \label{mccontour}}

\plotone{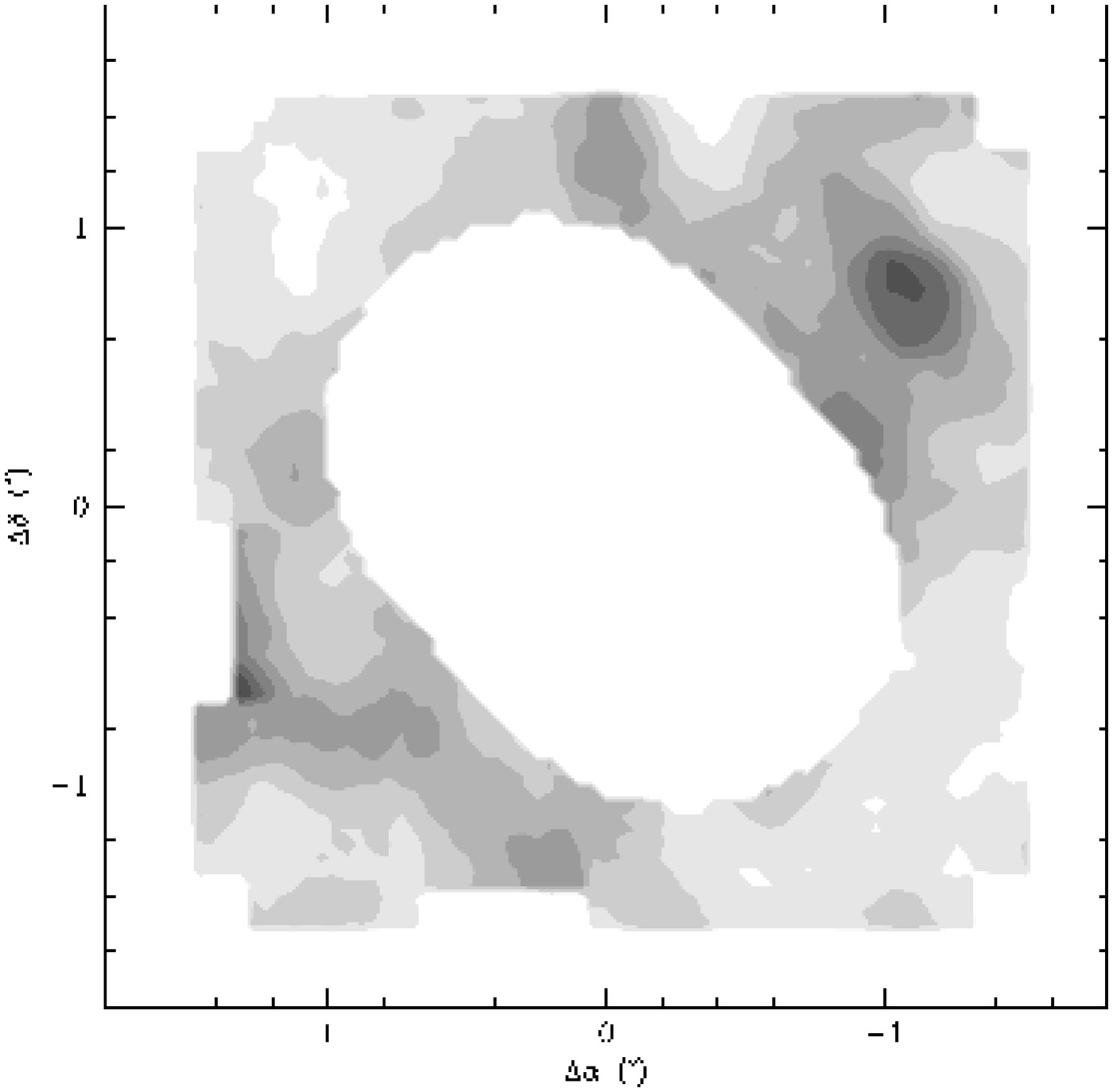}
\figcaption[Outer Contour Plot]{Filled contour plot with a smoothing scale of $24'$, where the grey-scale corresponds to the density bar in the distribution Fig.\ \ref{mccontour}.  The region inside the nominal tidal radius of Fornax is excluded.  The apparent enhancement located at $\Delta \alpha = 1.2^{\circ}$, $\Delta \delta  = -0.6^{\circ}$ is due to incompleteness effects at the edge of the survey area. \label{fornaxcontour_outer}}

\plotone{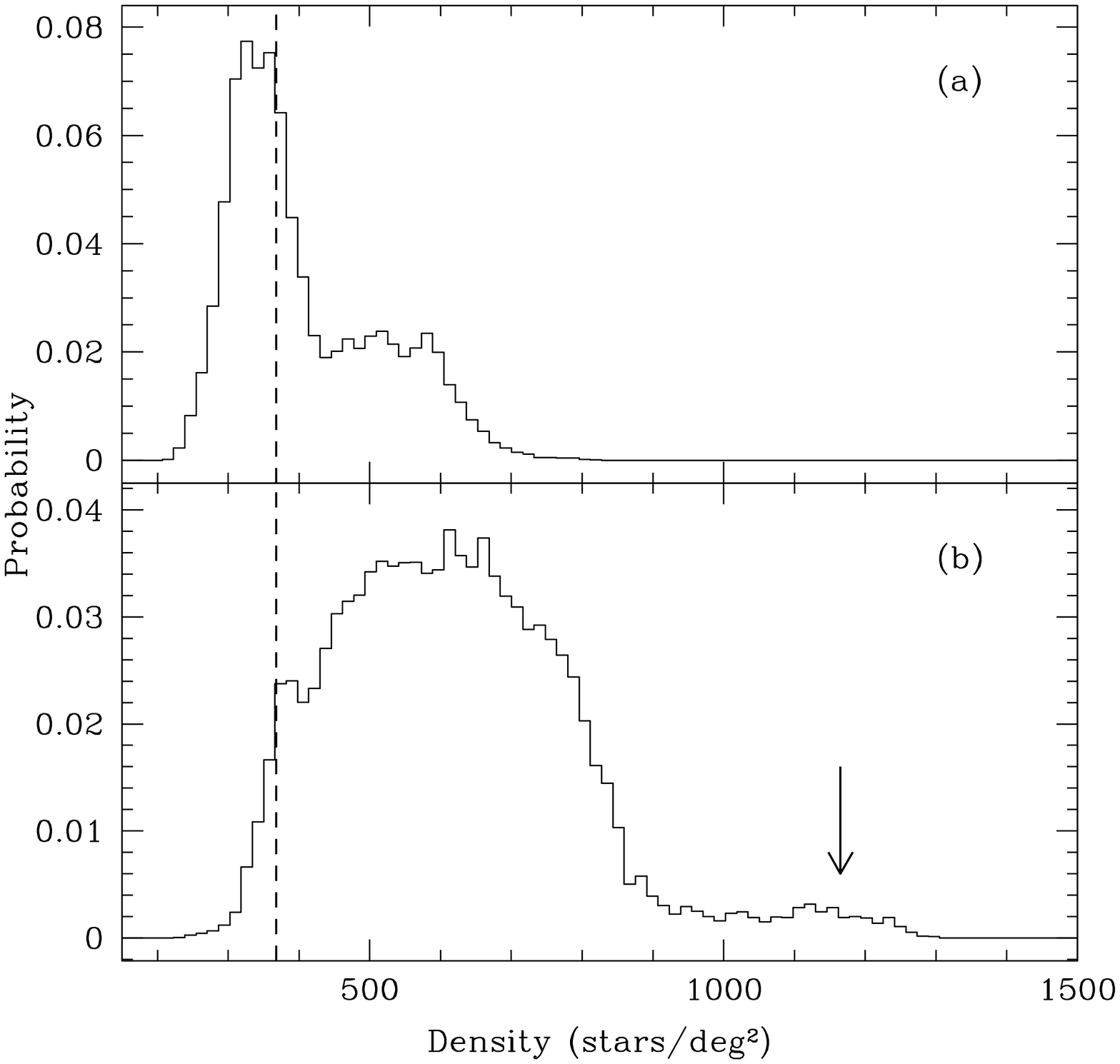}
\figcaption[Probability density function  -- four quadrants]{The probability density function evaluated separately for: (a) the field population-dominated regions to the NE and SW of Fornax, and, (b) the overdense NW/SE regions.  Both functions represent the density of stars beyond the nominal tidal radius.  The dashed line represents the normalised \citet{rat85} star counts, and the arrow in the lower panel indicates the high density peak from the shell-like structure in the NW quadrant.  \label{mcquads}}

\plotone{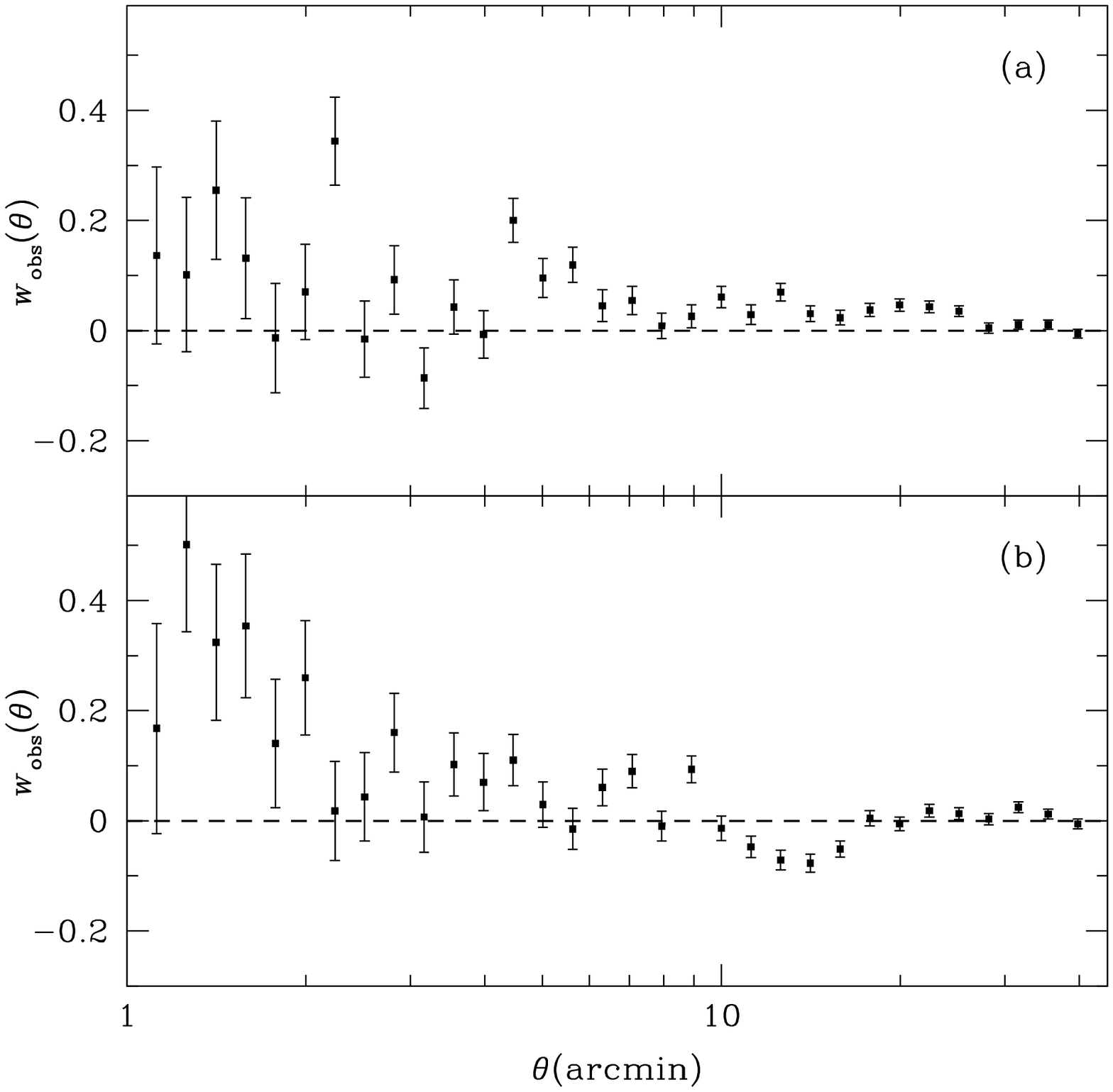}
\figcaption[Correlation Function]{(a) The {\em observed} angular correlation function for the NE section of the Fornax dataset.  The dashed line represents $w_{\mbox{\scriptsize obs}}(\theta) = 0$.  Error bars are measured from Poisson noise. (b) Same as (a) for the SW quadrant. \label{corrfn_boring}}

\plotone{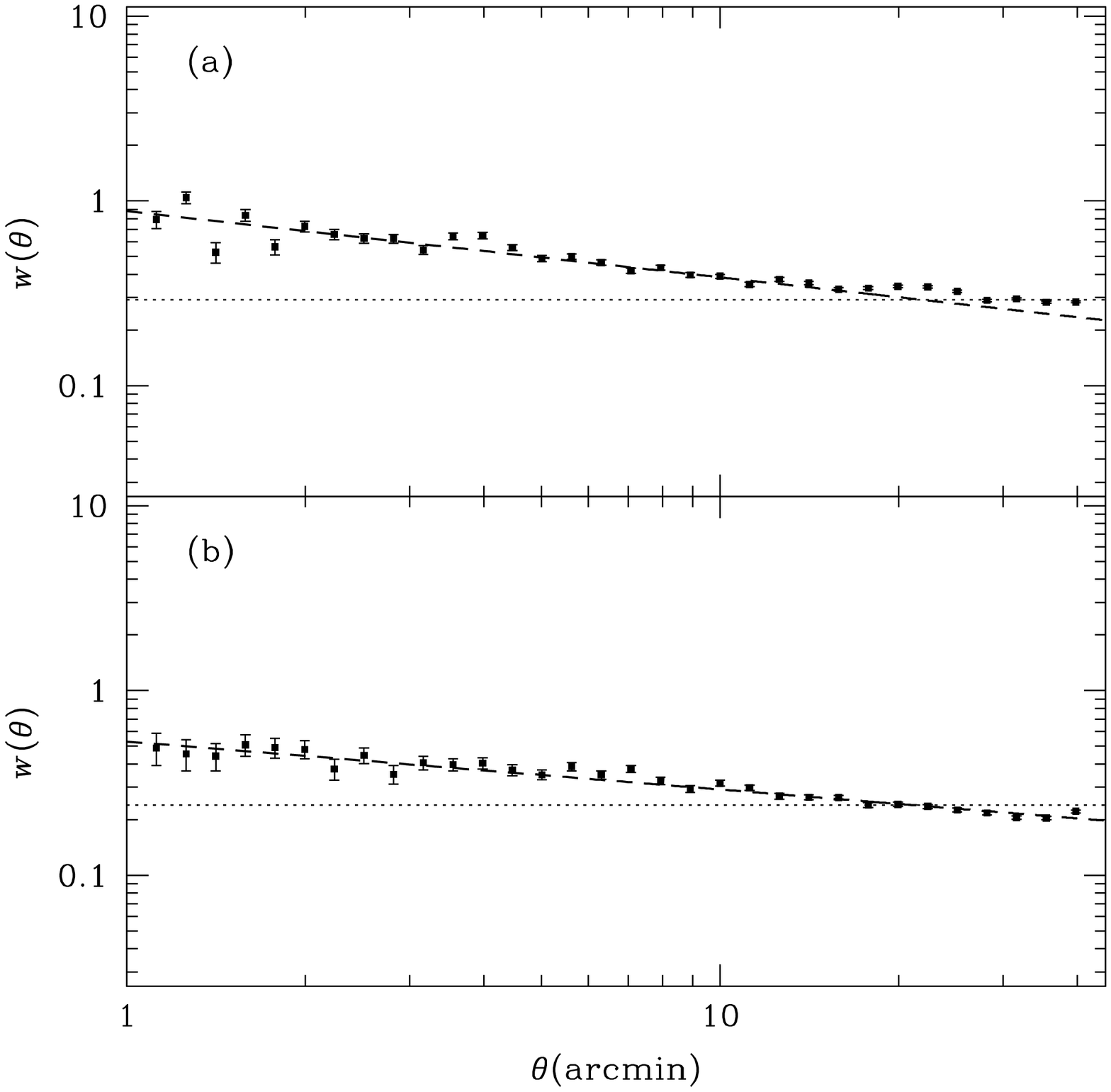}
\figcaption[Correlation Function]{(a) The angular correlation function measured for the extra-tidal NW quadrant.  Note that the vertical scale is logarithmic, as opposed to that in Fig.\ \ref{corrfn_boring}.  All points have been shifted by the integral constraint $A_w B$, and the dashed line represents the functional solution.  The dotted line represents $w(\theta)=A_w B$, equivalent to an observed angular correlation function of zero.  Error bars are measured from Poisson noise. (b) Same as (a) for the SE quadrant. \label{corrfn_interesting}}

\plotone{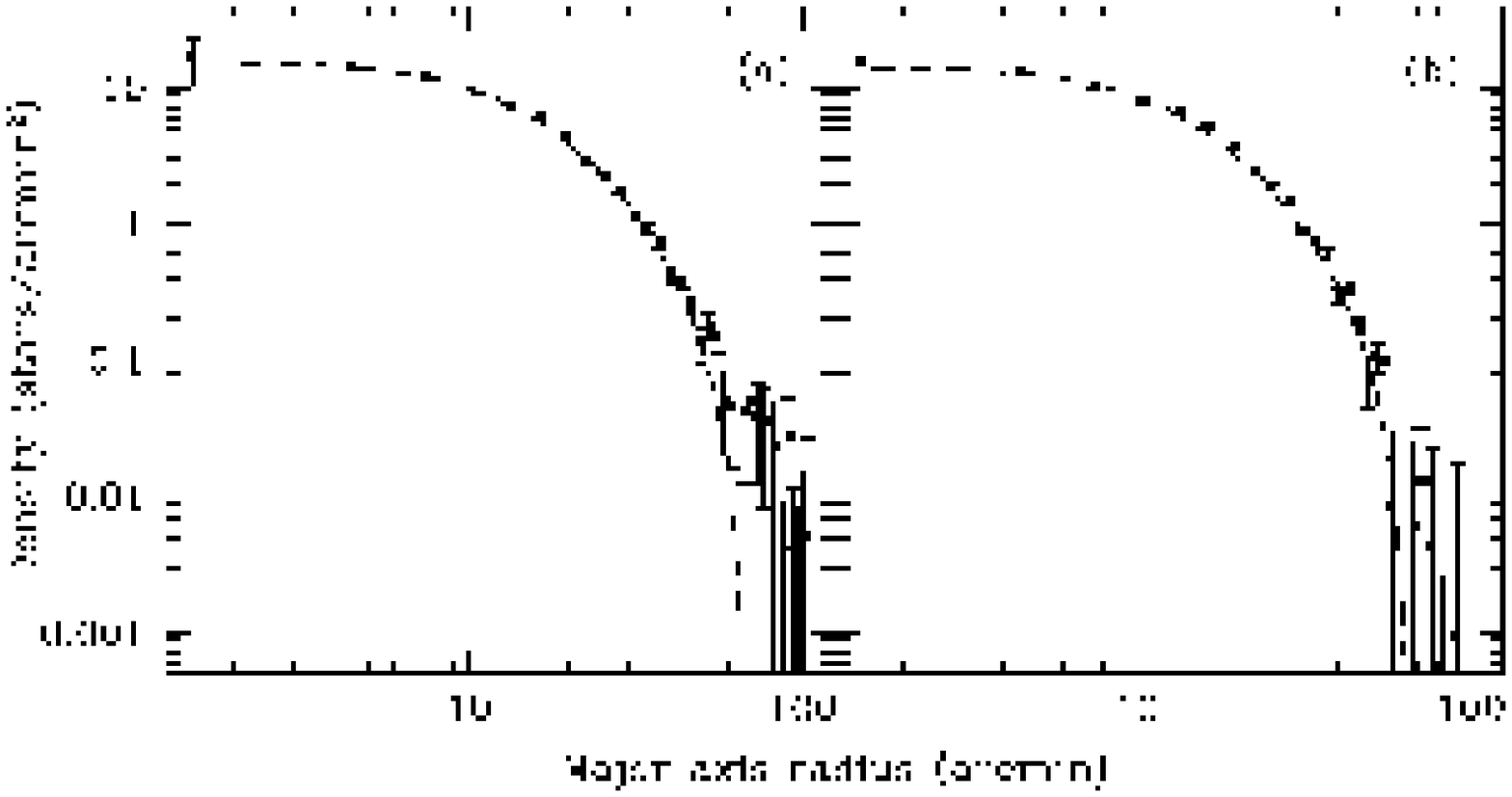}
\figcaption[Split radial profile of Fornax RGB stars]{Radial profiles for the two regions of Fornax; (a) NW/SE region, which contains the lobe structures beyond the tidal radius, and; (b) NE/SW region, with an extra-tidal region containing only the field population.  The background level (density of field stars) has been subtracted from each datapoint, and the dashed line represents the best-fitting King profile.  Both regions display approximately the same tidal and core radii as shown in Fig.\ \ref{radial}. \label{radialsplit}}

\plotone{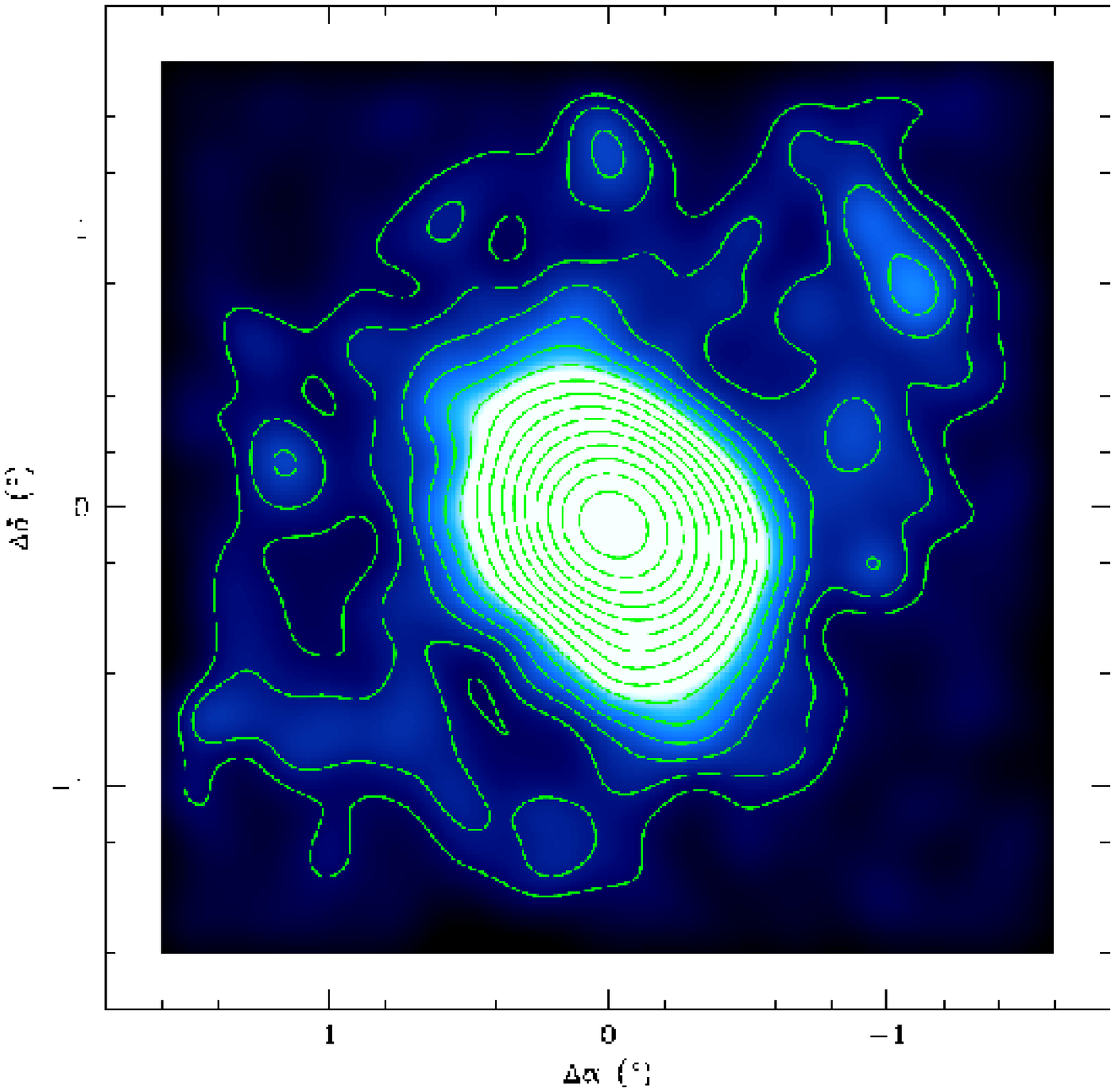}
\figcaption[Fornax Contour Plot]{Distribution of Fornax RGB stars where each star has been convolved with a Gaussian of radius $40''$.  The contours are logarithmically spaced, and are fitted with a smoothing length of $3.0'$.  The first contour represents a stellar density $3.3\sigma_{\mbox{\scriptsize field}}$ above the field star population.  \label{fornaxcontour}}

\plotone{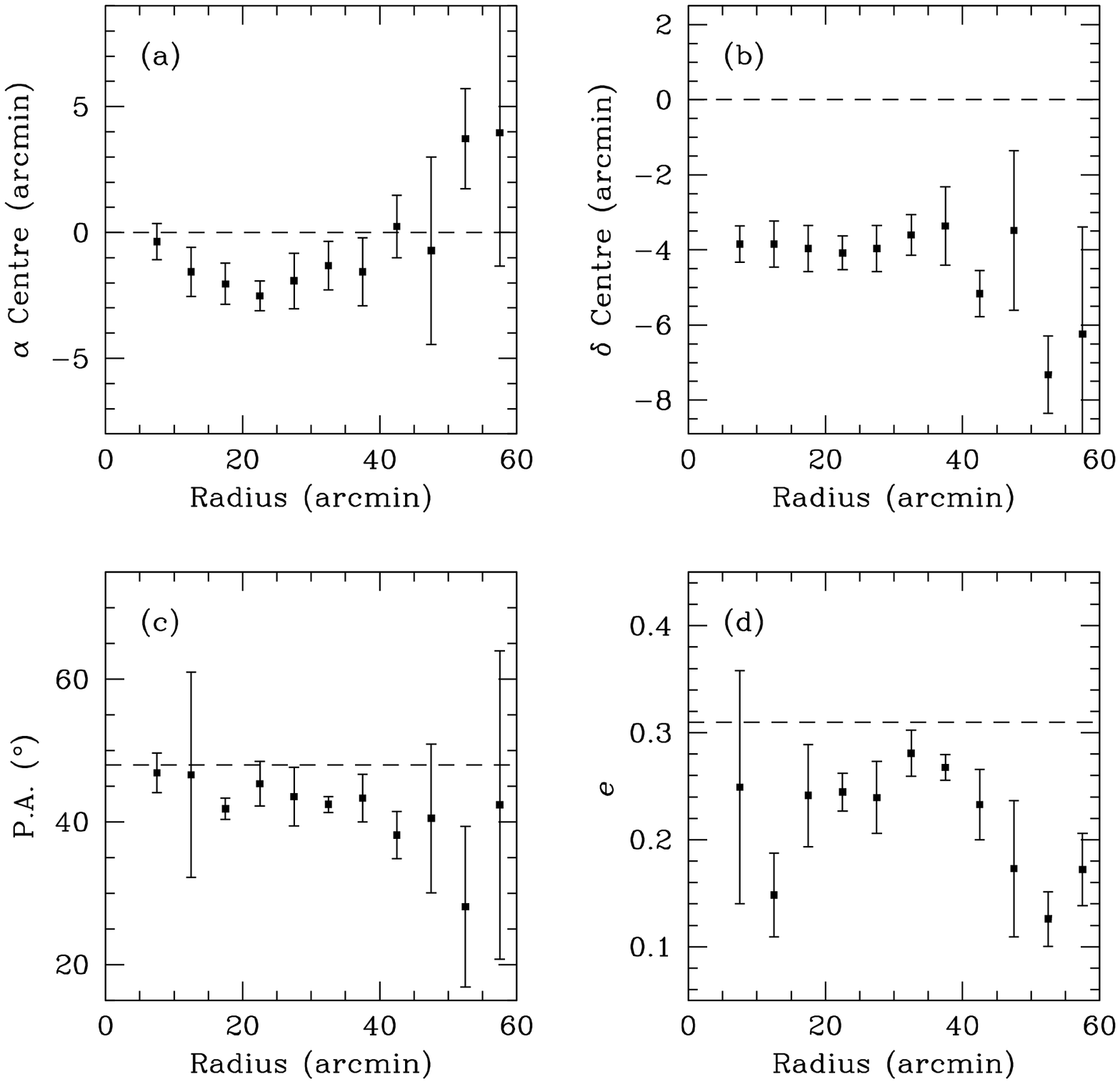}
\figcaption[Radial Dependence of Structure]{The dependence of Fornax structure on radius, where each point represents the binned result from five measurements at succeeding radii.  The dashed lines indicate the values listed in \citet{m98}.  Error bars are the standard deviations of each bin. \label{lscontour}}


\begin{sidewaystable}
\caption{List of Fornax observations}
\label{fornaxobservations}
\vspace{0.2cm}
\hspace{2cm}
\begin{tabular}{ccc|lcc|lcc}
\tableline
\tableline
 & & & & $V$ Images & & & $I$ Images & \\
Field & $\alpha$ (J2000.0) & $\delta$ (J2000.0) & Date Obs. & Exposures & Seeing & Date Obs. & Exposures & Seeing \\
\tableline
1 & 02:45:24.5 & -33:19:30 & 2002 Oct 30 & $8 \times 300$s &  $1.4''$ & 2002 Oct 30 &  $4 \times 480$s & $1.9''$ \\
2 & 02:41:47.6 & -33:19:30 & 2002 Nov 01 & $8 \times 300$s &  $1.6''$  & 2002 Nov 01 & $4 \times 480$s & $1.9''$ \\
3 & 02:38:10.4 & -33:19:30 & 2002 Nov 01 & $4 \times 600$s &  $1.7''$  & 2002 Nov 01 & $4 \times 480$s & $1.3''$ \\
4 & 02:34:33.3 & -33:19:30 & 2003 Dec 17 & $8 \times 300$s &  $2.0''$  & 2003 Dec 17 & $4 \times 480$s & $2.0''$ \\
5 & 02:45:24.5 & -34:04:30 & 2002 Nov 08 & $8 \times 300$s &  $1.4''$  & 2002 Nov 05 & $5 \times 480$s & $2.7''$ \\
6 & 02:41:47.6 & -34:04:30 & 2002 Oct 28 & $4 \times 600$s &  $1.7''$  & 2002 Oct 28 & $4 \times 480$s & $1.5''$ \\
7 & 02:38:10.4 & -34:04:30 & 2003 Dec 19 & $10 \times 300$s &  $1.6''$  & 2002 Oct 29 & $8 \times 240$s & $1.7''$ \\
8 & 02:34:33.3 & -34:04:30 & 2002 Nov 08 & $8 \times 300$s &  $1.8''$  & 2002 Nov 05 & $5 \times 480$s & $3.0''$ \\
9 & 02:45:03.5 & -34:53:47 & 2001 Oct 22 & $3 \times 600$s &  $2.0''$  & 2001 Oct 22 & $3 \times 400$s & $1.8''$ \\
10a & 02:41:34.2 & -34:48:07 & 2001 Oct 20 & $3 \times 600$s &  $2.2''$  & 2001 Oct 20 & $3 \times 400$s & $2.3''$ \\
10b & 02:41:34.2 & -34:48:07 & 2002 Oct 30 & $6 \times 600$s &  $1.8''$  & 2002 Nov 01 & $6 \times 480$s & $1.4''$ \\
11 & 02:38:10.4 & -34:49:30 & 2002 Oct 29 & $4 \times 600$s &  $1.4''$  & 2002 Oct 29 & $8 \times 240$s & $1.7''$ \\
12 & 02:34:33.3 & -34:49:30 & 2002 Nov 08 & $8 \times 300$s &  $1.8''$  & 2002 Nov 08 & $5 \times 480$s & $3.2''$ \\
13 & 02:45:24.5 & -35:34:30 & 2003 Dec 17 & $4 \times 600$s &  $2.1''$  & 2003 Dec 17 & $4 \times 480$s & $1.5''$ \\
14 & 02:41:33.4 & -35:25:24 & 2001 Oct 21 & $3 \times 600$s &  $1.7''$  & 2001 Oct 21 & $3 \times 400$s & $1.4''$ \\
15 & 02:38:10.4 & -35:34:30 & 2002 Nov 06 & $5 \times 600$s &  $2.9''$  & 2002 Nov 06 & $5 \times 480$s & $2.6''$ \\
16 & 02:34:33.3 & -35:34:30 & 2003 Dec 18 & $8 \times 300$s &  $1.9''$  & 2003 Dec 18 & $4 \times 480$s & $1.7''$ \\
\tableline
Total & --- & --- & --- & $41400$s &  $2.0''$  & --- & $30960$s & $1.8''$ \\
\tableline
\end{tabular}
\end{sidewaystable}

\begin{table}
\centering
\caption[Calibration values for WFI CCDs]{CCD calibration values over the WFI array.}
\label{ccdcalib}
\vspace{0.2cm}
\begin{tabular}{c|cc|cc}
\tableline
\tableline
CCD & $Z_V$ & $\Delta Z_V$ & $Z_I$ & $\Delta Z_I$ \\
\tableline
1 & $0.708 \pm 0.006$ & $-0.028 \pm 0.012$ & $1.156 \pm 0.005$ & $-0.085 \pm 0.009$ \\
2 & $0.736 \pm 0.010$ & --- & $1.241 \pm 0.007$ & --- \\
3 & $0.711 \pm 0.006$ & $-0.024 \pm 0.012$ & $1.239 \pm 0.005$ & $-0.002 \pm 0.009$ \\
4 & $0.749 \pm 0.014$ & $0.013 \pm 0.017$ & $1.277 \pm 0.013$ & $0.036 \pm 0.015$ \\
5 & $0.727 \pm 0.010$ & $-0.008 \pm 0.014$ & $1.226 \pm 0.006$ & $-0.016 \pm 0.009$ \\
6 & $0.736 \pm 0.010$ & $0.001 \pm 0.014$ & $1.262 \pm 0.007$ & $0.021 \pm 0.010$ \\
7 & $0.773 \pm 0.005$ & $0.038 \pm 0.011$ & $1.247 \pm 0.006$ & $0.006 \pm 0.009$ \\
8 & $0.719 \pm 0.011$ & $-0.016 \pm 0.015$ & $1.192 \pm 0.012$ & $-0.050 \pm 0.014$ \\
\tableline
\end{tabular}
\end{table}

\begin{table}
\centering
\caption[Magnitude conversions to match Field 10]{Magnitude shifts to adjust all magnitudes to the Field 10 photometric system.}
\label{fieldconv}
\vspace{0.2cm}
\begin{tabular}{c|cc}
\tableline
\tableline
Field & $\Delta Z_V$ & $\Delta Z_I$ \\
\tableline
1 & $0.035$ & $-0.161$ \\
2 & $0.039$ & $-0.074$ \\
3 & $-1.064$ & $-0.083$ \\
4 & $-0.322$ & $0.358$ \\
5 & $-0.061$ & $-0.022$ \\
6 & $0.050$ & $-0.149$ \\
7 & $-0.467$ & $-0.089$ \\
8 & $-0.916$ & $0.070$ \\
9 & $0.248$ & $0.186$ \\ 
10 & --- & --- \\
11 & $-0.691$ & $0.003$ \\
12 & $-1.001$ & $0.139$ \\
13 & $0.357$ & $0.352$ \\
14 & $0.263$ & $0.197$ \\
15 & $-0.765$ & $0.096$ \\
16 & $-0.443$ & $0.409$ \\
\tableline
\end{tabular}
\end{table}


\begin{thebibliography}{}
\bibitem[Aaronson \& Mould(1980)]{aaron80} Aaronson, M.~\& 
Mould, J.\ 1980, \apj, 240, 804
\bibitem[Aaronson \& Mould(1985)]{aaron85} Aaronson, M.~\& 
Mould, J.\ 1985, \apj, 290, 191 
\bibitem[Armandroff, Olszewski, \& Pryor(1995)]{arm95} 
Armandroff, T.~E., Olszewski, E.~W., \& Pryor, C.\ 1995, AJ, 110, 2131 
\bibitem[Bahcall \& Soneira(1980)]{bahcall80} Bahcall, J.~N.~\& 
Soneira, R.~M.\ 1980, \apjs, 44, 73 
\bibitem[Baugh et al.(1996)]{baugh96} 
Baugh, C.~M., Gardner, J.~P., Frenk, C.~S., \& Sharples, R.~M.\ 1996, 
\mnras, 283, L15 
\bibitem[Buonanno et al.(1998)]{buon98} Buonanno, R., Corsi, C. E., Zinn, R.,
Fusi Pecci, F., Hardy, E., \& Suntzeff, N. B. 1998, \apj, 501, L33
\bibitem[Buonanno et al.(1999)]{buon99} Buonanno, R., Corsi, C. E., 
Castellani, M., Marconi, G., Fusi Pecci, F., \& Zinn, R. 1999, \aj, 118, 1671
\bibitem[Carrera et al.(2002)]{carrera02} Carrera, R., Aparicio, A., 
Mart{\'{\i}}nez-Delgado, D., \& Alonso-Garc{\'{\i}}a, J.\ 2002, \aj, 123, 
3199
\bibitem[Coleman \& Da Costa(2005)]{coleman05} Coleman, M.~G.~\& Da Costa, G.~S.\ 2005, Publications of the Astronomical Society of Australia, 22, accepted
\bibitem[Coleman et al.(2004)]{coleman04a} Coleman, M., Da Costa, G.~S., Bland-Hawthorn, J., Mart\'{\i}nez-Delgado, D., Freeman, K.~C., \& Malin, D.\ 2004, AJ, 127, 832
\bibitem[Dekel, Devor, \& Hetzroni(2003)]{dekel03} Dekel, A., 
Devor, J., \& Hetzroni, G.\ 2003, \mnras, 341, 326
\bibitem[Demers et al.(1995)]{demers95} 
Demers, S., Battinelli, P., Irwin, M.~J., \& Kunkel, W.~E.\ 1995, \mnras, 
274, 491
\bibitem[Demers \& Irwin(1987)]{demers87} Demers, S.~\& Irwin, 
M.~J.\ 1987, \mnras, 226, 943
\bibitem[Demers, Irwin, \& Kunkel(1994)]{demers94} Demers, S., 
Irwin, M.~J., \& Kunkel, W.~E.\ 1994, \aj, 108, 1648 
\bibitem[Demers, Krautter, \& Kunkel(1980)]{demers80} Demers, 
S., Krautter, A., \& Kunkel, W.~E.\ 1980, \aj, 85, 1587 
\bibitem[Demers \& Kunkel(1979)]{demers79} Demers, S.~\& Kunkel, 
W.~E.\ 1979, \pasp, 91, 761
\bibitem[de Vaucouleurs \& Ables(1968)]{dev68} de 
Vaucouleurs, G.~\& Ables, H.~D.\ 1968, \apj, 151, 105 
\bibitem[Dinescu et al.(2004)]{dinescu04} Dinescu, D.~I., Keeney, B.~A., Majewski, S.~R., \& Girard, T.~M.\ 2004, AJ, 128, 687
\bibitem[Eskridge(1988a)]{eskridge88a} Eskridge, P.~B.\ 1988, \aj, 
96, 1352 
\bibitem[Eskridge(1988b)]{eskridge88b} Eskridge, P.~B.\ 1988, \aj, 
96, 1614
\bibitem[Freeman \& Bland-Hawthorn(2002)]{freeman02} Freeman, 
K.~\& Bland-Hawthorn, J.\ 2002, \araa, 40, 487 
\bibitem[Gallart et al.(1999)]{gallart99} Gallart, C., Freedman, 
W.~L., Aparicio, A., Bertelli, G., \& Chiosi, C.\ 1999, \aj, 118, 2245
\bibitem[G{\' o}mez-Flechoso \& Mart{\'{\i}}nez-Delgado(2003)]{spick03} G{\' o}mez-Flechoso, 
M.~{\' A}.~\& Mart{\'{\i}}nez-Delgado, D.\ 2003, ApJL, 586, L123 
\bibitem[Graham(1982)]{graham82} Graham, J.~A.\ 1982, \pasp, 94, 244 
\bibitem[Grebel, Gallagher, \& Harbeck(2003)]{grebel03} Grebel, 
E.~K., Gallagher, J.~S., \& Harbeck, D.\ 2003, \aj, 125, 1926 
\bibitem[Grillmair et al.(1995)]{grill95} 
Grillmair, C.~J., Freeman, K.~C., Irwin, M., \& Quinn, P.~J.\ 1995, \aj, 
109, 2553 
\bibitem[Hargreaves et al.(1994)]{har94} 
Hargreaves, J.~C., Gilmore, G., Irwin, M.~J., \& Carter, D.\ 1994, MNRAS, 
271, 693 
\bibitem[Helmi \& White(2001)]{helmi01} Helmi, A.~\& White, 
S.~D.~M.\ 2001, MNRAS, 323, 529 
\bibitem[Hernquist \& Quinn(1988)]{hernquist88} Hernquist, L.~\& Quinn, P.~J.\ 1988, ApJ, 331, 682
\bibitem[Hernquist \& Quinn(1989)]{hernquist89} Hernquist, L.~\& Quinn, P.~J.\ 1989, ApJ, 342, 1
\bibitem[Hewett(1982)]{hewett82} Hewett, P.~C.\ 1982, \mnras, 
201, 867 
\bibitem[Hodge(1961a)]{hodge61a} Hodge, P.~W.\ 1961, \aj, 66, 83
\bibitem[Hodge(1961b)]{hodge61b} Hodge, P.~W.\ 1961, \aj, 66, 249 
\bibitem[Hodge \& Smith(1974)]{hodge74} Hodge, P.~W.~\& Smith, 
D.~W.\ 1974, \apj, 188, 19
\bibitem[Ibata, Gilmore, \& Irwin(1994)]{ibata94} Ibata, R.~A., Gilmore, G., \& Irwin, M.~J.\ 1994, Nature, 370, 194 
\bibitem[Ibata et al.(2001)]{ibata01} Ibata, R., Lewis, G.~F., 
Irwin, M., Totten, E., \& Quinn, T.\ 2001, \apj, 551, 294 
\bibitem[Irwin \& Hatzidimitriou(1995)]{ih95} Irwin, M. \& Hatzidimitriou, D. 
1995, \mnras, 277, 1354
\bibitem[Johnston, Choi, \& Guhathakurta(2002)]{johnston02} 
Johnston, K.~V., Choi, P.~I., \& Guhathakurta, P.\ 2002, \aj, 124, 127
\bibitem[Johnston et al.(1999)]{johnston99} 
Johnston, K.~V., Zhao, H., Spergel, D.~N., \& Hernquist, L.\ 1999, \apjl, 
512, L109 
\bibitem[Kim et al.(2002)]{kim02} Kim, 
Y., Demarque, P., Yi, S.~K., \& Alexander, D.~R.\ 2002, \apjs, 143, 499 
\bibitem[King(1962)]{king62} King, I.\ 1962, \aj, 67, 471
\bibitem[King(1966)]{king66} King, I.~R.\ 1966, \aj, 71, 64 
\bibitem[Kleyna et al.(1998)]{kleyna98} Kleyna, J.~T., Geller, 
M.~J., Kenyon, S.~J., Kurtz, M.~J., \& Thorstensen, J.~R.\ 1998, \aj, 115, 
2359 
\bibitem[Kleyna et al.(2001)]{kleyna01} 
Kleyna, J.~T., Wilkinson, M.~I., Evans, N.~W., \& Gilmore, G.\ 2001, \apjl, 
563, L115 
\bibitem[Kleyna et al.(2003)]{kleyna03} Kleyna, J. T., Wilkinson, M. I.,
Gilmore, G., \& Evans, N. W. 2003, ApJ, 588, L21
\bibitem[Knapp, Kerr, \& Bowers(1978)]{knapp78} Knapp, G.~R., 
Kerr, F.~J., \& Bowers, P.~F.\ 1978, \aj, 83, 360 
\bibitem[Knebe et al.(2004)]{knebe04} Knebe, A., Gill, S.~P.~D., Kawata, D., \ Gibson, B.~K.\ 2004, MNRAS, submitted (astro-ph/0407418)
\bibitem[Landolt(1992)]{landolt92} Landolt, A.~U.\ 1992, \aj, 104, 340 
\bibitem[Landy \& Szalay(1993)]{landy93} Landy, S.~D.~\& 
Szalay, A.~S.\ 1993, \apj, 412, 64 
\bibitem[Law, Johnston, \& Majewski(2004)Law et al.]{law04} Law, D.~R., Johnston, K.~V., \& Majewski, S.~R.\ 2004, ApJ, submitted (astro-ph/0407566)
\bibitem[Limber(1954)]{limber54} Limber, D.~N.\ 1954, \apj, 119, 
655
\bibitem[Liske et al.(2003)]{liske03} Liske, J., Lemon, D.~J., 
Driver, S.~P., Cross, N.~J.~G., \& Couch, W.~J.\ 2003, \mnras, 344, 307 
\bibitem[Mackey \& Gilmore(2003)]{mackey03} Mackey, A.~D.~\& 
Gilmore, G.~F.\ 2003, \mnras, 340, 175
\bibitem[Maddox, Efstathiou, \& Sutherland(1996)]{maddox96} 
Maddox, S.~J., Efstathiou, G., \& Sutherland, W.~J.\ 1996, \mnras, 283, 
1227
\bibitem[Majewski et al.(2002)]{majewski02} Majewski, S.~R., et 
al.\ 2002, ASP Conf.~Ser.~285: Modes of Star Formation and the Origin of 
Field Populations, 199 
\bibitem[Majewski et al.(2003)]{majewski03} Majewski, S.~R., Skrutskie, M.~F., 
Weinberg, M.~D., \& Ostheimer, J.~C.\ 2003, \apj, 599, 1082 
\bibitem[Malin \& Carter(1980)]{malin80} Malin, D.~F.~\& Carter, D.\ 1980, Nature, 285, 643
\bibitem[Malin \& Carter(1983)]{malin83} Malin, D.~F.~\& 
Carter, D.\ 1983, \apj, 274, 534 
\bibitem[Malin, Quinn, \& Graham(1983)]{malin83b} Malin, D.~F., 
Quinn, P.~J., \& Graham, J.~A.\ 1983, \apjl, 272, L5
\bibitem[Mart{\'{\i}}nez-Delgado, Gallart, \& 
Aparicio(1999)]{mart-del99} Mart{\'{\i}}nez-Delgado, D., Gallart, 
C., \& Aparicio, A.\ 1999, \aj, 118, 862
\bibitem[Mart\'{\i}nez-Delgado et al.(2001)]{mart-del01} Mart\'{\i}nez-Delgado, D., Alonso-Garc\'{\i}a, J., Aparicio, A., \& G\'{o}mez-Flechoso, M. A.
2001, \apj, 549, L63
\bibitem[Mateo et al.(1991)]{mateo91} Mateo, M., Olszewski, E., 
Welch, D.~L., Fischer, P., \& Kunkel, W.\ 1991, \aj, 102, 914 
\bibitem[Mateo(1997)]{mateo97} Mateo, M.\ 1997, ASP 
Conf.~Ser.~116: The Nature of Elliptical Galaxies; 2nd Stromlo Symposium, 
259 
\bibitem[Mateo(1998)]{m98} Mateo, M. 1998, \araa, 36, 435
\bibitem[Mayer et al.(2001)]{mayer01} Mayer, L., Governato, F., 
Colpi, M., Moore, B., Quinn, T., Wadsley, J., Stadel, J., \& Lake, G.\ 
2001, ApJ, 559, 754 
\bibitem[McGaugh \& Bothun(1990)]{mcgaugh90} McGaugh, S.~S.~\& 
Bothun, G.~D.\ 1990, \aj, 100, 1073
\bibitem[Merrifield \& Kuijken(1998)]{merrifield98} Merrifield, 
M.~R.~\& Kuijken, K.\ 1998, \mnras, 297, 1292
\bibitem[Mighell \& Burke(1999)]{mighell99} Mighell, K.~J.~\& 
Burke, C.~J.\ 1999, \aj, 118, 366
\bibitem[Mould \& Aaronson(1983)]{mould83} Mould, J.~\& 
Aaronson, M.\ 1983, \apj, 273, 530 
\bibitem[Navarro, Abadi, \& Steinmetz(2004)]{navarro04} Navarro, 
J.~F., Abadi, M.~G., \& Steinmetz, M.\ 2004, \apjl, 613, L41
\bibitem[Olszewski \& Aaronson(1985)]{olsz85} Olszewski, 
E.~W.~\& Aaronson, M.\ 1985, \aj, 90, 2221
\bibitem[Odenkirchen et al.(2001)]{odenkirchen01} Odenkirchen, M., et 
al.\ 2001, \apjl, 548, L165 
\bibitem[Odenkirchen et al.(2003)]{odenkirchen03} Odenkirchen, M., et 
al.\ 2003, \aj, 126, 2385
\bibitem[Palma et al.(2003)]{palma03} Palma, C., Majewski, 
S.~R., Siegel, M.~H., Patterson, R.~J., Ostheimer, J.~C., \& Link, R.\ 
2003, AJ, 125, 1352
\bibitem[Peebles(1980)]{peebles80} Peebles, P.~J.~E.\ 1980, 
Research supported by the National Science Foundation.~Princeton, N.J., 
Princeton University Press, 1980.~435 p.,
\bibitem[Pence(1986)]{pence86} Pence, W.~D.\ 1986, \apj, 310, 597
\bibitem[Piatek et al.(2001)]{piatek01} Piatek, S., Pryor, C., Armandroff, 
T.~E., \& Olszewski, E.~W.\ 2001, \aj, 121, 841 
\bibitem[Pont et al.(2004)]{pont04} Pont, F., Zinn, R., 
Gallart, C., Hardy, E., \& Winnick, R.\ 2004, \aj, 127, 840
\bibitem[Ratnatunga \& Bahcall(1985)]{rat85} Ratnatunga, 
K.~U.~\& Bahcall, J.~N.\ 1985, \apjs, 59, 63 
\bibitem[Ricotti \& Wilkinson(2004)]{ricotti04} Ricotti, M.~\& Wilkinson, M.~I.\ 2004, \mnras, accepted, astro-ph/0406297
\bibitem[Roche \& Eales(1999)]{roche99} Roche, N.~\& Eales, 
S.~A.\ 1999, \mnras, 307, 703 
\bibitem[Rodgers \& Roberts(1994)]{rodgers94} Rodgers, A.~W.~\& 
Roberts, W.~H.\ 1994, \aj, 107, 1737
\bibitem[Sackett et al.(1994)]{sackett94} 
Sackett, P.~D., Rix, H., Jarvis, B.~J., \& Freeman, K.~C.\ 1994, \apj, 436, 
629 
\bibitem[Saviane, Held, \& Bertelli(2000)]{saviane00} Saviane, I., Held, E. V., \& Bertelli, G.\ 2000, \aap, 355, 56
\bibitem[Schiminovich et al.(1994)]{schim94} Schiminovich, D., van Gorkom, J.~H., van 
der Hulst, J.~M., \& Kasow, S.\ 1994, \apjl, 423, L101
\bibitem[Schlegel, Finkbeiner, \& Davis(1998)]{schlegel98} 
Schlegel, D.~J., Finkbeiner, D.~P., \& Davis, M.\ 1998, \apj, 500, 525 
\bibitem[Schweizer(1980)]{schweizer80} Schweizer, F.\ 1980, \apj, 
237, 303
\bibitem[Shapley(1939)]{shapley39} Shapley, H.\ 1939, Proceedings 
of the National Academy of Science, 25, 565
\bibitem[Stetson(1987)]{stetson87} Stetson, P. B. 1987, \pasp, 99, 191
\bibitem[Stetson(2000)]{stetson00} Stetson, P.~B.\ 2000, \pasp, 
112, 925
\bibitem[Stetson, Hesser, \& Smecker-Hane(1998)]{stetson98} Stetson, P. B., 
Hesser, J. E., \& Smecker-Hane, T. A.\ 1998, \pasp, 110, 533 
\bibitem[Sung \& Bessell(2000)]{sung00} Sung, H.~\& Bessell, 
M.~S.\ 2000, Publications of the Astronomical Society of Australia , 17, 244
\bibitem[Susa \& Umemura(2004)]{susa04} Susa, H.~\& Umemura, 
M.\ 2004, \apj, 600, 1 
\bibitem[Tolstoy et al.(2001)]{tolstoy01} Tolstoy, E., Irwin, 
M.~J., Cole, A.~A., Pasquini, L., Gilmozzi, R., \& Gallagher, J.~S.\ 2001, 
\mnras, 327, 918
\bibitem[Tolstoy et al.(2003)]{tolstoy03} Tolstoy, E., Venn, K. A., Shetrone,
 M., Primas, F., Hill, V., Kaufer, A., \& Szeifert, T. 2003, \aj, 125, 707 
\bibitem[Tosi(2003)]{tosi03} Tosi, M. 2003, AP\&SS, 284, 651
\bibitem[Walcher et al.(2003)]{walcher03} Walcher, C.~J., Fried, J.~W., 
Burkert, A., \& Klessen, R.~S.\ 2003, \aap, 406, 847 
\bibitem[Yi et al.(2001)]{yi01} Yi, S., Demarque, P., Kim, 
Y., Lee, Y., Ree, C.~H., Lejeune, T., \& Barnes, S.\ 2001, \apjs, 136, 417 
\bibitem[Young(1999)]{young99} Young, L.~M.\ 1999, \aj, 117, 
1758 
\bibitem[Zacharias et al.(2000)]{zach00} Zacharias, N., et 
al.\ 2000, \aj, 120, 2131 
\end{thebibliography}
\end{document}